\documentclass[final,3p,times]{elsarticle}
\usepackage{amssymb}
\usepackage{amssymb}
\usepackage[]{amsmath}
\usepackage{graphics}
\usepackage{amssymb}
\usepackage[]{amsmath}
\usepackage{amsthm}
\usepackage{setspace}
\usepackage{epsfig}
\usepackage{subfigure}
\usepackage{cases}
\usepackage{graphicx}

\biboptions{numbers,sort&compress}

\usepackage{amsmath}
\usepackage{times}
\usepackage{anysize}
\marginsize{2.5cm}{2.5cm}{1cm}{2cm}

\selectfont

 \journal {}

\begin{document}

\begin{frontmatter}

\title{Localized excitations and interactional solutions for the reduced Maxwell-Bloch equations}

\author{ Lili Huang$^{1,2}$ }
\author{ Yong Chen$^{1,2,3}$ \corref{cor1} }
\ead{ychen@sei.ecnu.edu.cn}

\cortext[cor1]{Corresponding author. ******}

\address[label1]{Shanghai Key Laboratory of Trustworthy Computing, East China Normal University,
    Shanghai, 200062,  China}

\address[label2]{MOE International Joint Lab of Trustworthy Software, East China Normal University,
    Shanghai, 200062, China}

\address[label3]{Department of Physics, Zhejiang Normal University,
    Jinhua, 321004,  China}

\begin{abstract}
Based on nonlocal symmetry method, localized excitations and interactional solutions are investigated for the reduced Maxwell-Bloch equations. The nonlocal symmetries of the reduced Maxwell-Bloch equations are obtained by the truncated Painlev\'{e} expansion approach and the M\"{o}bious invariant property. The nonlocal symmetries are localized to a prolonged system by introducing suitable auxiliary dependent variables. The extended system can be closed and a novel Lie point symmetry system is constructed. By solving the initial value problems, a new type of finite symmetry transformations is obtained to derive periodic waves, Ma breathers and breathers travelling on the background of periodic line waves. Then rich exact interactional solutions are derived between solitary waves and other waves including cnoidal waves, rational waves, Painlev\'{e} waves, and periodic waves through similarity reductions. In particular, several new types of localized excitations including rogue waves are found, which stem from the arbitrary function generated in the process of similarity reduction. By computer numerical simulation, the dynamics of these localized excitations and interactional solutions are discussed, which exhibit meaningful structures.
\end{abstract}

\end{frontmatter}

\section{Introduction}
Maxwell equations are foundation and core of the electromagnetic theory. It was first presented at `A Dynamical Theory of the Electromagnetic Field', published on the Royal Society by British physicist James Clerk Maxwell \cite{j-maxwelljc-ptrsl-1865} in 1865, based on Coulomb's law, Biot-Savart law, and Faraday's law. Bloch equations are one of the most important theoretical foundations of mechanical characterization on nuclear magnetic resonance, and also the basis of studying on the coherent optical transient phenomena. The Bloch equations were introduced by Felix Bloch \cite{j-blochf-pr-1946} in 1946. It is well known that the Maxwell-Bloch equations successfully demonstrate the propagation of short ($<10^{-9}s$) optical pulses in resonant media \cite{m-maimistovai-1999}. In the dielectric of two level atoms, the most basic semiclassical equations governing the propagation of Electromagnetic waves are the Maxwell wave equation
\begin{equation}\label{rmb-001}
E'_{\xi\xi}(\xi,\tau)-c^{-2}E'_{\tau\tau}(\xi,\tau)=\frac{4\pi}{c^2}np<u_{\tau\tau}(\xi,\tau,\mu')>,
\end{equation}
and the Bloch type equations
\begin{eqnarray}\label{rmb-002}
&& u_{\tau}(\xi,\tau,\mu')=-\mu' v(\xi,\tau,\mu'),\notag\\
&& v_{\tau}(\xi,\tau,\mu')=\mu' u(\xi,\tau,\mu')+2p\hbar^{-1}E'(\xi,\tau)\omega(\xi,\tau,\mu'),\notag\\
&& \omega_{\tau}(\xi,\tau,\mu')=-2p\hbar^{-1}E'(\xi,\tau)v(\xi,\tau,\mu').
\end{eqnarray}

The system of equations (\ref{rmb-001}) and (\ref{rmb-002}) is the Maxwell-Bloch (MB) equations \cite{j-ejcbrk-jpagp-1972} with $E'(\xi,\tau)$ the electric field, $u(\xi,\tau,\mu')$ the microscopic polarization, $v(\xi,\tau,\mu')$ the phase information, $\omega(\xi,\tau,\mu')$ the atomic inversion, $\mu'$ the proportional of atomic resonant frequency, $p$ the matrix element to the dipole operator, $n$ the atomic dipole density, c the speed of light in vacuum. The angular bracket $<>$ represents the summation over all the media demonstrated by the frequency, and $\hbar\mu'$ is the energy separation of the two levels. The subscripts $\xi$ and $\tau$ refer to partial differentiation with respect to the space and time.

In the low density approximation, the Maxwell equation (\ref{rmb-001}) can be reduced to an equation describing waves travelling only to the right \cite{j-eilbeckjc-jpa-1972}
\begin{equation}\label{rmb-003}
E'_\xi+c^{-1}E'_\tau=-2\pi c^{-1}np<u_\tau>.
\end{equation}
Eq. (\ref{rmb-003}) is a good approximation at atomic densities. The system of Eqs. (\ref{rmb-002}) and (\ref{rmb-003}) are called the reduced Maxwell-Bloch (RMB) equations \cite{j-eilbecklc-jpa-1973}. The propagation of ultra-short optical pulses is usually governed by the RMB equations.

In 1973, Eilbeck et al \cite{j-eilbecklc-jpa-1973} first began the research of the RMB equations, in which slowly varying envelope approximation was avoided and the backscattered wave was neglected as a weaker assumption. Then the RMB equations enjoy a wide usage in describing phenomena in nonlinear optics, namely the theory of optical self-induced transparency (SIT). A type of coherent optical soliton in a two-level resonant system related with the SIT phenomenon was first put forward by McCall and Hahn \cite{j-mccallsl-prl-1967}. The RMB equations represent the propagation of a short laser pulse in a rarefied medium of two level atoms. The RMB equations are important in the physics of nonlinear optics and have been observed in a lot of experiments \cite{m-jens-1978} with essential features. With the tremendous progress of laser technology, ultra-short optical pulses have attracted much more attentions \cite{j-brabect-pmp-2000,j-wazwaz-amc-2005,j-sunb-jars-2012,j-xusw-pre-2013,j-fengbf-pre-2016}.

For plane polarized waves, the RMB equations are found integrable and connected with a Zakharov-Shabat scattering problem. Many effective methods, such as the inverse scattering transform (IST) \cite{m-mjablowitz-1991,j-gibbonjd-lnc-1973}, the Hirota bilinear method \cite{m-rhirota-2004,j-caudreypj-jam-1974,j-caudreypj-ptrsa-2011}, the Darboux transformation (DT) \cite{m-vbmatveev-1991,j-guor-nd-2012,j-shigh-sm-2017}, the Painlev\'{e} analysis \cite{j-xugq-cpc-2009}, etc., have been developed to study the explicit N-soliton solutions of the RMB equations. The RMB equations are one of integrable systems as shown in \cite{j-grauela-jpa-1986,j-grauela-lnc-1984} and of course admit other integrable properties including Hamitonian structure and recursion operator \cite{j-aiyerrn-jam-1983}, the N-degenerate periodic solutions, N-rational solutions and rogue waves \cite{j-weij-arxive-2017} as well as interactional solutions \cite{j-huangll-cpb-2018} by consistent Riccati expansion method. Owing to the practical significance, further study of the characteristics of the RMB equations, especially the interactional solutions, is therefore of great importance.


For a long time, the symmetry theory play an important role in solving nonlinear systems, whether integrable system or not. To find interactional solutions of nonlinear systems is a difficult and tedious but very important and meaningful work. Fortunately, recent studies \cite{j-huxr-pre-2012,j-sylou-jpa-2012,j-chengxp-pre-2014,j-chenjc-jmp-2014,j-huangll-aml-2017} have shown that nonlocal symmetry method is one of the best tools to find nonlinear waves interacting with each other. Since nonlocal symmetries were first researched by Vinogradov and Krasil'shchik \cite{j-vinogradovam-dans-1980} early in 1980, many mathematicians \cite{j-akhatovis-jms-1991,m-blumangw-2010,j-galasf-jpa-1992,j-guthriega-jpa-1993,j-sylou-jpa-1997,j-sylou-jpamg-1997} have made great contributions to the development of the research on nonlocal symmetry. In recent years, nonlocal symmetry method achieves further progress in solving nonlinear system. In \cite{j-huxr-pre-2012,j-sylou-jpa-2012}, the nonlocal symmetries, which related to the DT and B\"{a}cklund transformation (BT), were investigated to derive soliton-cnoidal interactional wave solutions. In order to avoid missing some important results such as integral terms or high order derivative terms of nonlocal variables in the symmetries, a systematic method \cite{j-xinxp-cpl-2013} was put forward to find the nonlocal symmetries of nonlinear systems. Lately, it comes to light that nonlocal symmetries can be derived by Painlev\'{e} analysis \cite{j-sylou-arxiv-2013}. As related to the truncated Painlev\'{e} expansion, this kind of nonlocal symmetries is just the residual of the expansion in regard to singular manifold, and is also known as residual symmetry \cite{j-gaoxn-jhep-2013,j-renbo-nd-2016,j-huangll-cpb-2016,j-renb-cnsns-2017}. So far, nonlocal symmetries of the RMB equations have not been reported. To our knowledge, localized excitations such as rogue waves and breathers of the nonlinear systems have not yet been found by the nonlocal symmetry method up to the present.

Rogue waves \cite{j-peregrinedh-jamssb-1983,j-akhmedievn-pre-2009,j-bludovyv-epjst-2010,j-guobl-pre-2012,j-linhlm-pre-2013,j-zhaolc-pre-2016,j-chenjc-pla-2015,j-wangx-cnsns-2015} have become a hot topic in plenty of research areas in recent years. The concept of freak rogue wave was proposed by Draper \cite{j-draperl-mo-1965} in 1965 in the ocean. Rogue waves are localized in both space and time, which appear from nowhere and disappear without a trace \cite{j-akhmedievn-pla-2009}, have taken responsibility for numerous marine disasters. At present, the most complete recorded rogue wave was the `new year's wave' \cite{j-walkerdag-aor-2005}, spotted in Jan. 1, 1995 in North sea. The optical rogue wave \cite{j-sollidr-nature-2007} was first discovered by experiments in nonlinear optics in 2007. The peregrine soliton \cite{j-kibler-np-2010} in nonlinear fibre optics was observed in 2010. Up to now, the investigation of rogue waves have occurred in many areas, such as atmosphere \cite{j-stenflo-jpp-2010}, capillary flow \cite{j-shatsm-prl-2010}, Bose-Einstein condensates \cite{j-bludovyv-pra-2009}, superfluid \cite{j-efimovvb-epjst-2010}, and even finance \cite{j-yanzy-ctp-2010}. Breathers \cite{j-tajirim-pre-1999,j-kedzioradj-pre-2012,j-hejs-pre-2013,j-liuc-pra-2014} are regarded as the crucial prototypes to explain rogue wave phenomena and are the localized breathing waves with a periodic profile in a certain direction. In many of the existing literatures mainly reported two types of breathers such as Akhmediev breathers \cite{j-akhmedievnn-tmp-1986,j-akhmedievn-pra-2009} and Ma breathers \cite{j-mayc-sam-1979}, the former are space-periodic breather solutions and the latter are time-periodic breather solutions. In real life, there are amount of waves interacting with each other. In the process of interaction, many localized excitations phenomena will appear. The solitons are the most basic excitations of the integrable systems. In 1988, the dromion solution of the DS system was obtained by Boiti et al \cite{j-boiti-pla-1988}. Since then, plenty of works \cite{j-fokasas-pd-1990,j-hietarintaj-pla-1990,j-radhar-jnmp-1999} have been focus on the localized excitations of the nonlinear integrable systems. Lately, variable separation solutions of the integrable systems were derived by the variable separation method \cite{j-tangxy-pre-2002,j-tangxy-jmp-2003,j-lousy-jpa-2003}. Because some arbitrary functions were included in the variable separation solutions, abundant localized excitations, such as solitoff solutions, dromion solutions, breathers, instantons, and ghost solitons were obtained. The richness of localized excitations may play a significant contribution to investigate dynamical properties of the nonlinear models.

In this paper, the RMB equations are investigated by nonlocal symmetry method. We focus on constructing localized excitations and interactional solutions of the RMB equations. The nonlocal symmetries of the RMB equations are obtained by the truncated Painlev\'{e} expansion approach and the M\"{o}bious invariant property. By introducing suitable auxiliary dependent variables, the nonlocal symmetries are localized to a prolonged system. The prolonged system can be closed to a Lie point symmetry and a new type of finite symmetry transformations is derived by solving the initial value problems. It is difficult to find new explicit solutions stemming from nonconstant nonlinear wave such as the cnoidal waves and Painlev\'{e} waves. Based on the finite symmetry transformations, we obtain periodic waves, Ma breathers and breathers travelling on the background of periodic line waves. By symmetry reduction method, rich exact interactional solutions are derived between solitary waves and other waves, such as cnoidal waves, rational solutions, Painlev\'{e} waves, and periodic waves. As an arbitrary function generated during the similarity reduction process, several new types of localized excitations including rogue waves, breathers and other nonlinear waves are obtained. In particular, the dynamics of these interactional solutions are discussed and shown to exhibit meaningful structures by computer numerical simulation.

This paper is arranged as follows. 
In Section 2, nonlocal symmetries of the RMB equations are obtained from the truncated Painlev\'{e} expansion approach. Then the nonlocal symmetries are localized to Lie point symmetries by prolonging the original system to a large system. In section 3, the finite symmetry transformations and similar reductions of the prolonged system are presented to derive preiodic wave solutions, breathers, interactional solutions and rogue waves of the original system. The last section contains a short summary and discussion.

\section{Nonlocal symmetry and its localization}

Through the following transformation
\begin{equation}\label{rmb-005}
x=\tau-c^{-1}\xi,~~t=-4\pi np^2(c\hbar)^{-1}\mu\xi,~~E=2p\hbar^{-1}E',
\end{equation}
sends the RMB equations to
\begin{eqnarray}\label{rmb-03}
&& u_x=-\mu v,~~v_x=E\omega+\mu u,\notag\\
&& \omega_x=-Ev,~~E_t=-v.
\end{eqnarray}

As the RMB equations (\ref{rmb-03}) pass the Painlev\'{e} test  \cite{j-grauela-jpa-1986}, they  have the truncated Painlev\'{e} expansion
\begin{eqnarray}\label{rmb-04}
&& u=u_0+\frac{u_1}{\phi},~~v=v_0+\frac{v_1}{\phi}+\frac{v_2}{\phi^2},\notag\\
&& \omega=\omega_0+\frac{\omega_1}{\phi}+\frac{\omega_2}{\phi^2},~~E=E_0+\frac{E_1}{\phi},
\end{eqnarray}
with $u_0,u_1,v_0,v_1,v_2,\omega_0,\omega_1,\omega_2,E_0,E_1,$ and $\phi$ being the function of $x$ and $t$. The manifold $\phi(x,t)=0$ is often called a movable singularity manifold.

To find nonlocal symmetries of the RMB equations (\ref{rmb-03}) related to their truncated Painlev\'{e} expansion, one has to substitute (\ref{rmb-04}) into (\ref{rmb-03}) and balance all the coefficients of different powers of $\phi$,
\begin{eqnarray}\label{rmb-05}
&& u_0=-I\mu\phi_x^{-1}\phi_{xt},~~u_1=2I\mu\phi_t,~~v_0=I\phi_x^{-1}\phi_{xxt}-I\phi_x^{-2}\phi_{xx}\phi_{xt},\notag\\
&&v_1=-2I\phi_{xt},~~v_2=2I\phi_x\phi_t,~~\omega_0=-\mu^2\phi_x^{-1}\phi_t-\phi_x^{-1}\phi_{xxt}+\phi_x^{-2}\phi_{xx}\phi_{xt},\\
&&\omega_1=2\phi_{xt},~~\omega_2=-2\phi_x\phi_t,~~E_0=-I\phi_x^{-1}\phi_{xx},~~E_1=2I\phi_x,\notag
\end{eqnarray}
and the RMB equations (\ref{rmb-03}) are successfully reduced to the following Schwarzian equation:
\begin{equation}\label{rmb-06}
\mu^2C_x+S_t=0,
\end{equation}
where $C=\frac{\phi_t}{\phi_x}$ is the general Schwarzian variable, $S=\frac{\phi_{xxx}}{\phi_x}-\frac{3}{2}\Big(\frac{\phi_{xx}}{\phi_x}\Big)^2$ is the Schwarzian derivative.

In accordance with the definition of residual symmetries, the nonlocal symmetries of the RMB equations (\ref{rmb-03}) can be read out from truncated Painlev\'{e} analysis (\ref{rmb-04})
\begin{equation}\label{rmb-08}
\sigma^u=2I\mu\phi_t,~~\sigma^v=-2I\phi_{xt},~~\sigma^\omega=2\phi_{xt},~~\sigma^E=2I\phi_x,
\end{equation}
It is necessary to point out that the Schwarzian form (\ref{rmb-06}) is invariant under the M\"{o}bious transformation
\begin{equation}\label{rmb-09}
\phi\rightarrow\frac{a+b\phi}{c+d\phi} ~~ (ad\neq bc),
\end{equation}
which means (\ref{rmb-06}) possesses the symmetry $\sigma^{\phi}=-\phi^2$ in a special case $a=0, b=c=1, d=\epsilon$. The nonlocal symmetries (\ref{rmb-08}) can also be derived by substituting the M\"{o}bious transformation symmetry $\sigma^\phi$ into the linearized equations of $u_0,v_0,\omega_0,$ and $E_0$ in (\ref{rmb-05}).

By introducing another three new dependent variables $f\equiv f(x,t),g\equiv g(x,t),$ and $h\equiv h(x,t)$, the nonlocal symmetries of the RMB equations (\ref{rmb-03}) can be localized to Lie point symmetries
\begin{equation}\label{rmb-12}
\sigma^u=2I\mu f,~~\sigma^v=-2Ig,~~\sigma^\omega=2g,~~\sigma^E=2Ih,~~\sigma^f=-2\phi f,~~\sigma^g=-2(\phi g+fh),~~\sigma^h=-2\phi h,~~\sigma^\phi=-\phi^2,
\end{equation}
for the prolonged system
\begin{eqnarray}\label{rmb-13}
\nonumber && u_x+\mu v=0,~~v_x-E\omega-\mu u=0,~~\omega_x+Ev=0,~~E_t+v=0,~~u=-I\mu\phi_x^{-1}\phi_{xt},~~f=\phi_t,\\
&& v=I\phi_x^{-1}\phi_{xxt}-I\phi_x^{-2}\phi_{xx}\phi_{xt},~~
\omega=-\mu^2\phi_x^{-1}\phi_t-\phi_x^{-1}\phi_{xxt}+\phi_x^{-2}\phi_{xx}\phi_{xt},~~E=-I\phi_x^{-1}\phi_{xx},\\
\nonumber && g=f_x,~~h=\phi_x,~~\mu^2\phi_x^2\phi_{xt}-\mu^2\phi_x\phi_{xx}\phi_t+\phi_x^2\phi_{xxxt}-3\phi_x\phi_{xx}\phi_{xxt}
-\phi_x\phi_{xxx}\phi_{xt}+3\phi_{xx}^2\phi_{xt}=0.
\end{eqnarray}
Finally, the prolonged system (\ref{rmb-13}) can be closed with the vector form
\begin{equation}\label{rmb-14}
V=2I\mu f\frac{\partial}{\partial u}-2Ig\frac{\partial}{\partial v}+2g\frac{\partial}{\partial \omega}+2Ih\frac{\partial}{\partial E}-2\phi f\frac{\partial}{\partial f}-2(\phi g+fh)
\frac{\partial}{\partial g}-2\phi h\frac{\partial}{\partial h}-\phi^2\frac{\partial}{\partial \phi}.
\end{equation}

\section{Explicit solutions from nonlocal symmetry}
After making the nonlocal symmetries (\ref{rmb-08}) equivalent to Lie point symmetry (\ref{rmb-14}), one can construct the explicit solutions by the Lie group theory in two ways.

\textbf{A. Finite symmetry transformation}

Due to Lie point symmetry (\ref{rmb-14}), by means of solving the initial value problem:
\begin{eqnarray}
\nonumber   &&\frac{d\bar{u}(\epsilon)}{d\epsilon}=2I\mu\bar{f},~~\frac{d\bar{v}(\epsilon)}{d\epsilon}=-2I\bar{g},
~~\frac{d\bar{\omega}(\epsilon)}{d\epsilon}=2\bar{g},~~\frac{d\bar{E}(\epsilon)}{d\epsilon}=2I\bar{h},\\
\label{rmb-15} &&\frac{d\bar{f}(\epsilon)}{d\epsilon}=-2\bar{\phi}\bar{f},~~\frac{d\bar{g}(\epsilon)}{d\epsilon}=-2(\bar{f}\bar{h}+\bar{\phi}\bar{g}),~~\frac{d\bar
{h}(\epsilon)}{d\epsilon}=-2\bar{\phi}\bar{h},~~\frac{d\bar{\phi}(\epsilon)}{d\epsilon}=-\bar{\phi}^2,\\
\nonumber  &&\bar{u}(0)=u,~~\bar{v}(0)=v,~~\bar{\omega}(0)=\omega,~~\bar{E}(0)=E,~~\bar{f}(0)=f,~~\bar{g}(0)=g,~~\bar{h}(0)=h,~~\bar{\phi}(0)=\phi,
\end{eqnarray}
where $\epsilon$ is group parameter, we can obtain the following symmetry group theorem:\\
\textbf{Theorem 1.} If $\{u,v,\omega,E,f,g,h,\phi\}$ is a solution of the extended system (\ref{rmb-13}), then so is $\{\bar{u},\bar{v},\bar{\omega},\bar{E},\bar{f},\bar{g},\bar{h},\bar{\phi}\}$ given by
\begin{eqnarray}
\nonumber  &&\bar{u}=u+\frac{2I\epsilon\mu f}{1+\epsilon \phi},~~\bar{v}=v-\frac{2I\epsilon g}{1+\epsilon \phi}+\frac{2I\epsilon^2 fh}{(1+\epsilon \phi)^2},~~\bar{\omega}=\omega+\frac{2\epsilon g}{1+\epsilon \phi}-\frac{2\epsilon^2 fh}{(1+\epsilon \phi)^2},\\
\label{rmb-16} &&\bar{E}=E+\frac{2I\epsilon h}{1+\epsilon \phi},~~\bar{f}=\frac{f}{(1+\epsilon \phi)^2},~~\bar{g}=\frac{g}{(1+\epsilon \phi)^2}-\frac{2\epsilon fh}{(1+\epsilon \phi)^3},~~\bar{h}=\frac{h}{(1+\epsilon \phi)^2},~~\bar{\phi}=\frac{\phi}{1+\epsilon \phi}.
\end{eqnarray}

\textbf{Remark 1.} For a known solution $\{u,v,\omega,E\}$ of (\ref{rmb-03}), the above finite symmetry transformation will derive another solution $\{\bar{u},\bar{v},\bar{\omega},\bar{E}\}$. It should be point out that the last equation of Eqs. (\ref{rmb-16}) is nothing but the corresponding known M\"{o}bious transformation.

In order to clearly illustrate Theorem 1, we consider the following three types of periodic solutions for the RMB equations (\ref{rmb-03}), which means that nontrivial solutions for the RMB equations can be derived from trivial solutions by using the finite symmetry transformations (\ref{rmb-16}):

\textbf{Type 1}: Considering the trivial solution $u=\mu l_1,v=0,\omega=-\frac{l_1\mu^2}{k_1}$, and $E=k_1$ of the RMB equations (\ref{rmb-03}), periodic solutions for the introduced dependent variables can be obtained:
\begin{equation}\label{rmb-j1}
f=Il_1\exp(I(k_1x+l_1t)),~~g=-k_1l_1\exp(I(k_1x+l_1t)),~~h=Ik_1\exp(I(k_1x+l_1t)),~~\phi=1+\exp(I(k_1x+l_1t)).
\end{equation}
Substituting (\ref{rmb-j1}) into (\ref{rmb-16}) yields nontrivial solutions of the RMB equations (\ref{rmb-03}):
\begin{eqnarray}
\nonumber  &&u=\mu l_1-\frac{2\epsilon\mu l_1\exp(I(k_1x+l_1t))}{1+\epsilon(1+\exp(I(k_1x+l_1t)))},\\
\nonumber  &&v=\frac{2I\epsilon k_1l_1\exp(I(k_1x+l_1t))}{1+\epsilon(1+\exp(I(k_1x+l_1t)))}-\frac{2I\epsilon^2k_1l_1\exp(2I(k_1x+l_1t))}{(1+\epsilon(1+\exp(I(k_1x+l_1t))))^2},\\
\label{rmb-j2}  &&\omega=-\frac{l_1\mu^2}{k_1}-\frac{2\epsilon k_1l_1\exp(I(k_1x+l_1t))}{1+\epsilon(1+\exp(I(k_1x+l_1t)))}+\frac{2\epsilon^2k_1l_1\exp(2I(k_1x+l_1t))}{(1+\epsilon(1+\exp(I(k_1x+l_1t))))^2},\\
\nonumber  &&E=k_1-\frac{2\epsilon k_1\exp(I(k_1x+l_1t))}{1+\epsilon(1+\exp(I(k_1x+l_1t)))}.
\end{eqnarray}

\begin{figure*}[!htbp]
\centering
\subfigure[]{\includegraphics[height=1.4in,width=1.9in]{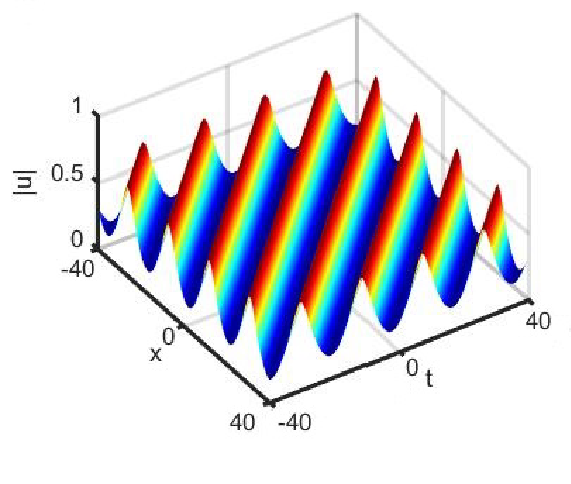}}\hspace{1cm}
\subfigure[]{\includegraphics[height=1.4in,width=1.9in]{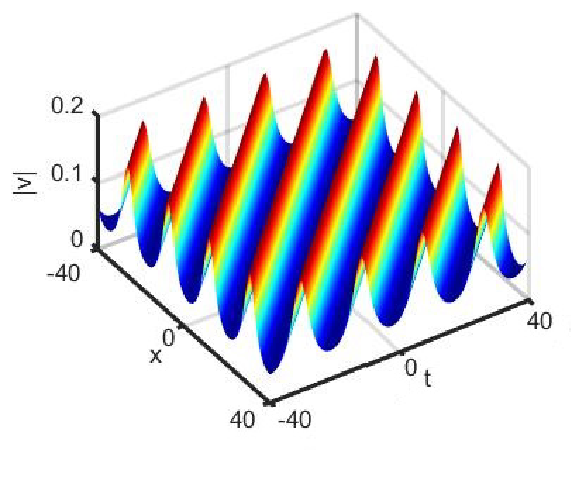}}\hspace{1cm}
\subfigure[]{\includegraphics[height=1.4in,width=1.9in]{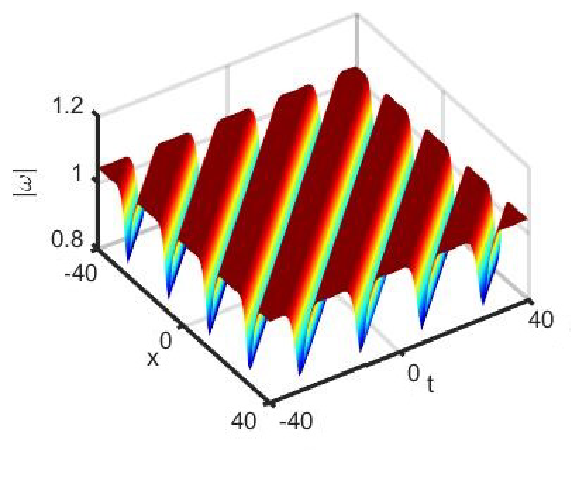}}\hspace{1cm}
\subfigure[]{\includegraphics[height=1.4in,width=1.9in]{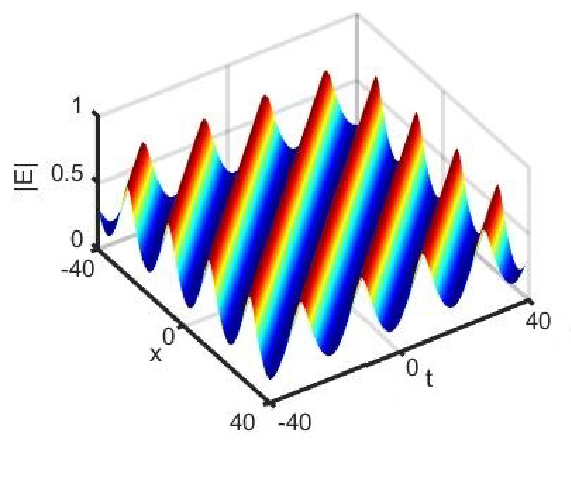}}
\caption{ The periodic wave solutions of the RMB equations for the components $u,v,\omega,$ and $E$ expressed by (\ref{rmb-j2}). The parameters are $\mu=1,k_1=\frac{1}{3},l_1=\frac{1}{3},\epsilon=\frac{1}{2}$.}\label{Fig-01}
\end{figure*}

The periodic wave solutions (\ref{rmb-j2}) are shown in Fig. (\ref{Fig-01}). It can be seen that these solutions are periodic in both $x$ and $t$, and maintain constant amplitudes. The maximum amplitudes of the components $|u|,|v|,|\omega|,$ and $|E|$ are $0.6666,0.1667,1.0438,$ and $0.6666$, respectively. If a kink-solitary wave solution for (\ref{rmb-06}) is chosen as $\phi=\tanh(I(k_2x+l_2t))$, solitary wave solutions of the RMB equations (\ref{rmb-03}) can also be obtained by using the finite symmetry transformations (\ref{rmb-16}).

\textbf{Type 2}: If we choose $\phi=1+\exp(I(k_1x+l_1t))+\exp(I(k_2x+l_2t))$ in (\ref{rmb-13}), then fundamental solutions can be derived with $k_1=k_2=I\mu$ and $l_1,l_2$ are arbitrary constants:
\begin{eqnarray}
\nonumber  &&v=0,~~E=I\mu,~~h=-\mu(\exp(-\mu x+Il_1t)+\exp(-\mu x+Il_2t)),\\
\nonumber  &&u=\mu(l_1\exp(-\mu x+Il_1t)+l_2\exp(-\mu x+Il_2t))(\exp(-\mu x+Il_1t)+\exp(-\mu x+Il_2t))^{-1},\\
\label{rmb-j3}  &&\omega=I\mu(l_1\exp(-\mu x+Il_1t)+l_2\exp(-\mu x+Il_2t))(\exp(-\mu x+Il_1t)+\exp(-\mu x+Il_2t))^{-1},\\
\nonumber  &&f=I(l_1\exp(-\mu x+Il_1t)+l_2\exp(-\mu x+Il_2t)),~~g=-I\mu((l_1\exp(-\mu x+Il_1t)+l_2\exp(-\mu x+Il_2t))).
\end{eqnarray}

\begin{figure*}[!htbp]
\centering
\subfigure[]{\includegraphics[height=1.4in,width=1.9in]{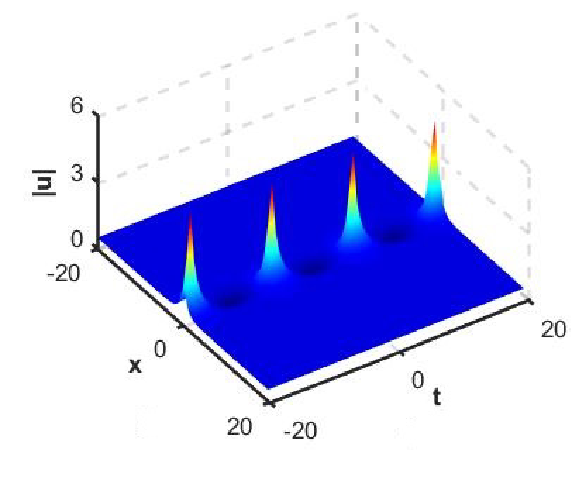}}\hspace{1cm}
\subfigure[]{\includegraphics[height=1.4in,width=1.9in]{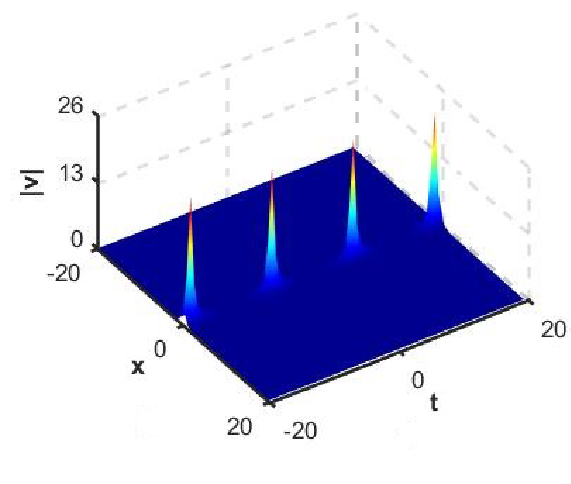}}\hspace{1cm}
\subfigure[]{\includegraphics[height=1.4in,width=1.9in]{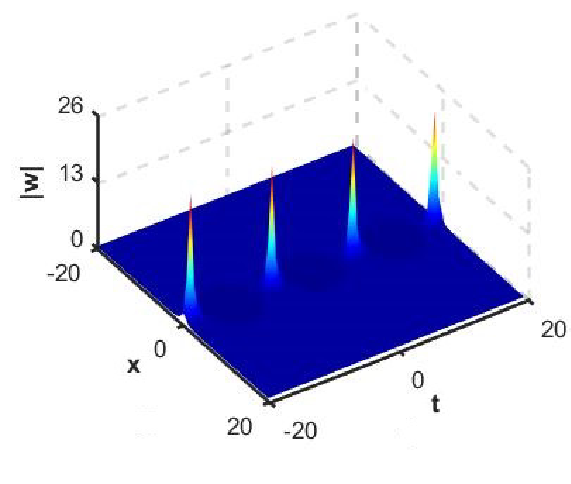}}\hspace{1cm}
\subfigure[]{\includegraphics[height=1.4in,width=1.9in]{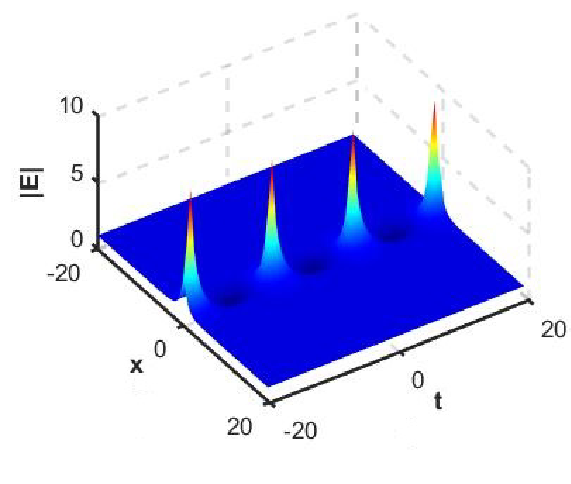}}
\caption{ The breathers of the RMB equations for the components $u,v,\omega,$ and $E$. The parameters are $\mu=1,l_1=\frac{1}{2},l_2=\frac{1}{2},\epsilon=1$.}\label{Fig-01-b}
\end{figure*}

Substituting (\ref{rmb-j3}) into (\ref{rmb-16}) yields breather solutions of the RMB equations (\ref{rmb-03}). The breathers for the components $u,v,\omega,$ and $E$ are shown in Fig. (\ref{Fig-01-b}). It can be seen that these solutions are periodic in $t$ directions and localized in $x$ directions, and maintain constant amplitudes, which are known as Ma breathers \cite{j-mayc-sam-1979}.

\textbf{Type 3}: If we choose $\phi=1+\exp(I(k_1x+l_1t))+\exp(I(k_2x+l_2t))+\exp(I(k_3x+l_3t))$ in (\ref{rmb-13}), then fundamental solutions can be derived with $k_1=k_2=k_3=I\mu$ and $l_1,l_2,l_3$ are arbitrary constants:
\begin{eqnarray}
\nonumber  &&u=\mu(l_1\exp(-\mu x+Il_1t)+l_2\exp(-\mu x+Il_2t)+l_3\exp(-\mu x+Il_3t))(\exp(-\mu x+Il_1t)\\
\nonumber  &&~~~~+\exp(-\mu x+Il_2t)+\exp(-\mu x+Il_3t))^{-1},~~v=0,\\
\nonumber  &&\omega=I\mu(l_1\exp(-\mu x+Il_1t)+l_2\exp(-\mu x+Il_2t)+l_3\exp(-\mu x+Il_3t))(\exp(-\mu x+Il_1t)\\
\label{rmb-j4}  &&~~~~+\exp(-\mu x+Il_2t)+\exp(-\mu x+Il_3t))^{-1},~~E=I\mu,\\
\nonumber  &&f=I(l_1\exp(-\mu x+Il_1t)+l_2\exp(-\mu x+Il_2t)+l_3\exp(-\mu x+Il_3t)),\\
\nonumber  &&g=-I\mu(l_1\exp(-\mu x+Il_1t)+l_2\exp(-\mu x+Il_2t)+l_3\exp(-\mu x+Il_3t)),\\
\nonumber  &&h=-\mu(\exp(-\mu x+Il_1t)+\exp(-\mu x+Il_2t)+\exp(-\mu x+Il_3t)).
\end{eqnarray}

\begin{figure*}[!htbp]
\centering
\subfigure[]{\includegraphics[height=1.4in,width=1.9in]{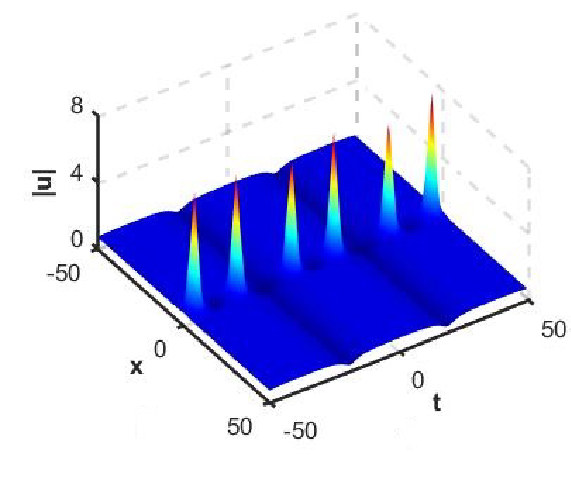}}\hspace{1cm}
\subfigure[]{\includegraphics[height=1.4in,width=1.9in]{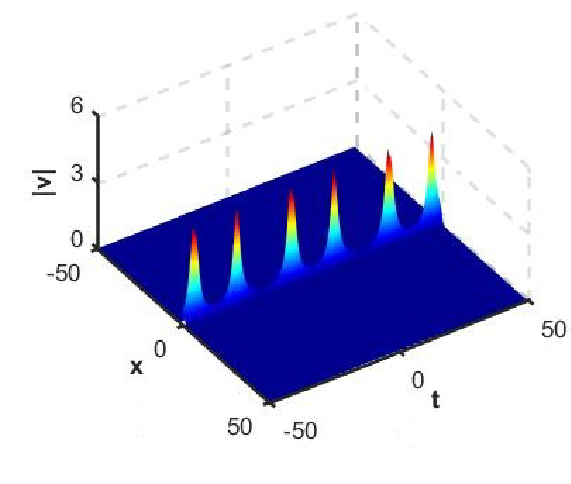}}\hspace{1cm}
\subfigure[]{\includegraphics[height=1.4in,width=1.9in]{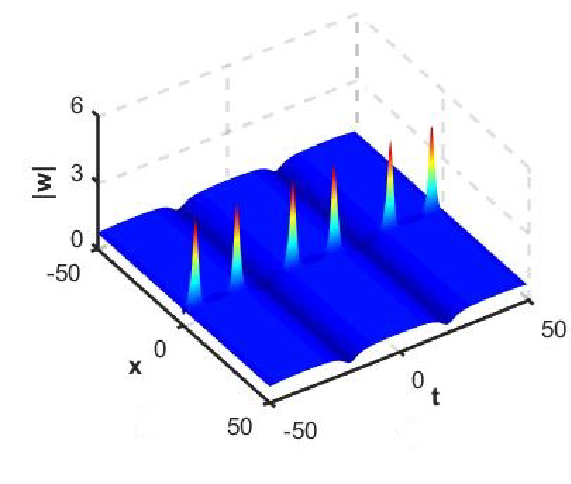}}\hspace{1cm}
\subfigure[]{\includegraphics[height=1.4in,width=1.9in]{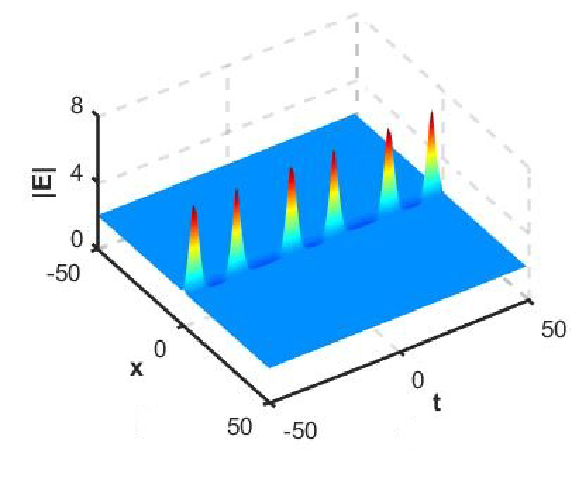}}
\caption{ The hybrid solutions of the RMB equations for the components $u,v,\omega,$ and $E$. The parameters are $\mu=2,l_1=\frac{1}{3},l_2=\frac{1}{3},l_3=\frac{1}{2},\epsilon=1$.}\label{Fig-02-b}
\end{figure*}

Substituting (\ref{rmb-j4}) into (\ref{rmb-16}) yields hybrid of breathers and periodic wave solutions for the RMB equations (\ref{rmb-03}). The hybrid solutions for the components $u,v,\omega,$ and $E$ are shown in Fig. (\ref{Fig-02-b}). As can be seen that the solutions for the components $u$ and $\omega$ describe breathers travelling on the background of periodic line waves. Both the breathers and periodic line waves are periodic in $t$ directions and localized in $x$ directions. The solutions for the components $v$ and $E$ describe breathers as in type 2 with different parameters. For the component $v$, they both come from zero plane wave backgrounds. For the component $E$, they both come from nonzero plane wave backgrounds. To our knowledge, it has not discovered this type of solutions for the RMB equations (\ref{rmb-03}).

\textbf{B. similarity reductions of the prolonged system}

In order to construct similarity reductions of (\ref{rmb-03}), we apply the Lie symmetry method to the whole extended system. Supposing Eqs. (\ref{rmb-13}) are invaraint under this infinitesimal transformations
\begin{equation}\label{rmb-17}
\{x,t,u,v,\omega,E,f,g,h,\phi\}\rightarrow\{x+\epsilon X,t+\epsilon T,u+\epsilon U,v+\epsilon V,\omega+\epsilon W,E+\epsilon M,f+\epsilon F,g+\epsilon G,h+\epsilon H,\phi+\epsilon\Phi\}
\end{equation}
with
\begin{eqnarray}\label{rmb-18}
&&\sigma^u=Xu_x+Tu_t-U,~~\sigma^v=Xv_x+Tv_t-V,~~\sigma^\omega=X\omega_x+T\omega_t-W,~~\sigma^E=XE_x+TE_t-M,\notag\\
&&\sigma^f=Xf_x+Tf_t-F,~~\sigma^g=Xg_x+Tg_t-G,~~\sigma^h=Xh_x+Th_t-H,~~\sigma^\phi=X\phi_x+T\phi_t-\Phi,
\end{eqnarray}
where $X,T,U,V,W,M,F,G,H$, and $\Phi$ are functions of $\{x,t,u,v,\omega,E,f,g,h,\phi\}$, and $\epsilon$ is a small parameter.

Next, substituting (\ref{rmb-18}) into the linearized symmetry equations for the extended system
\begin{eqnarray}\label{rmb-19}
&&\sigma^u_x+\mu\sigma^v=0,~~\sigma^v_x-\mu\sigma^u-E\sigma^{\omega}-\omega\sigma^E=0,~~\sigma^{\omega}_x+E\sigma^v+v\sigma^E=0,~~\sigma^E_t+\sigma^v=0,~~\sigma^f-\sigma^\phi_t=0,\notag\\
&&I\phi_x^2\sigma^u+\mu\phi_{xt}\sigma^{\phi}_x-\mu\phi_x\sigma^{\phi}_{xt}=0,~~I\phi_x^3\sigma^v-\phi_x\phi_{xxt}\sigma^{\phi}_x+2\phi_{xx}\phi_{xt}\sigma^{\phi}_x
-\phi_x\phi_{xt}\sigma^{\phi}_{xx}-\phi_x\phi_{xx}\sigma^{\phi}_{xt}+\phi_x^2\sigma^{\phi}_{xxt}=0,\notag\\
&&\phi_x^3\sigma^{\omega}+\mu^2\phi_x^2\sigma^{\phi}_t-(\mu^2\phi_x\phi_t+\phi_x\phi_{xxt}-2\phi_{xx}\phi_{xt})\sigma^{\phi}_{x}-
\phi_x\phi_{xt}\sigma^{\phi}_{xx}-\phi_x\phi_{xx}\sigma^{\phi}_{xt}+\phi_x^2\sigma^{\phi}_{xxt}=0,~~\sigma^g-\sigma^f_x=0,\\
&&(\mu^2\phi_x\phi_{xx}\phi_t+3\phi_x\phi_{xx}\phi_{xxt}-6\phi_{xx}^2\phi_{xt}+\phi_x\phi_{xxx}\phi_{xt})\sigma^{\phi}_x-\mu^2\phi_x^2\phi_{xx}\sigma^{\phi}_t-(
\mu^2\phi_x^2\phi_t-6\phi_x\phi_{xx}\phi_{xt}+3\phi_x^2\phi_{xxt})\sigma^{\phi}_{xx}\notag\\
&&+(\mu^2\phi_x^3+3\phi_x\phi_{xx}^2-\phi_x^2\phi_{xxx})\sigma^{\phi}_{xt}-3\phi_x^2
\phi_{xx}\sigma^{\phi}_{xxt}-\phi_x^2\phi_{xt}\sigma^{\phi}_{xxx}+\phi_x^3\sigma^{\phi}_{xxxt}=0,~~I\phi_x^2\sigma^{E}+\phi_{xx}\sigma^{\phi}_x-\phi_x\sigma^{\phi}_{xx}=0,\notag\\
&&\sigma^h-\sigma^{\phi}_x=0,\notag
\end{eqnarray}
Then collecting the coefficients of all the variables and their partial derivatives, and setting all of them to zero, we get a system of overdetermined linear equations with the infinitesimals $\{x,t,u,v,\omega,E,f,g,h,\phi\}$. After solving them, we can obtain
\begin{eqnarray}\label{rmb-20}
&&X=c_4,~~T=f_1,~~U=-Ic_1\mu f-f_{1t}u,~~V=Ic_1g-vf_{1t},~~W=-c_1g-f_{1t}\omega,~~M=-Ic_1h,\notag\\
&&F=c_1\phi f+c_2f-f_{1t}f,~~G=c_1(fh+\phi g)+c_2g-f_{1t}g,~~H=c_1\phi h+c_2h,~~\Phi=\frac{1}{2}c_1\phi^2+c_2\phi+c_3,
\end{eqnarray}
where $f_1\equiv f_1(t)$ is an arbitrary function of $t$, and $c_i(i=1\ldots4)$ are four arbitrary constants. Especially, when $c_2=c_3=c_4=f_1=0$ and $c_1=-2$, the derived symmetry is just Eq. (\ref{rmb-14}), and when $c_1=0$, the related symmetry is just the general Lie point symmetry of the original Eq. (\ref{rmb-03}).

For the sake of more corresponding group invariant solutions, the following characteristic equations need to be solved:
\begin{equation}\label{rmb-21}
\frac{dx}{X}=\frac{dt}{T}=\frac{du}{U}=\frac{dv}{V}=\frac{d\omega}{W}=\frac{dE}{M}=\frac{df}{F}=\frac{dg}{G}=\frac{dh}{H}
=\frac{d\phi}{\Phi}.
\end{equation}
Next, two nontrivial similarity reductions originated from (\ref{rmb-21}) will be detailed study under the condition $c_1\neq0$.

\textbf{Reduction 1.} Without loss of generality, we hypothesize $c_4=1$ and $f_1=k$, and define the parameter $c$ by $c^2=\frac{c_2^2-2c_1c_3}{4}$ $(c\neq0)$. By solving (\ref{rmb-21}), we obtain similarity solutions
\begin{eqnarray}\label{rmb-22}
&&u=U+\frac{c_1}{c}I\mu F\tanh\Delta_1,~~v=V+\frac{c_1}{c}IG\tanh\Delta_1+\frac{c_1^2}{2c^2}IFH\tanh^2\Delta_1,\notag\\
&&\omega=W-\frac{c_1}{c}G\tanh\Delta_1-\frac{c_1^2}{2c^2}FH\tanh^2\Delta_1,~~E=M+\frac{Ic_1}{c}H\tanh\Delta_1,~~h=-H{\rm sech}^2\Delta_1,\\
&&f=-F{\rm sech}^2\Delta_1,~~g=(G+\frac{c_1}{c}FH\tanh\Delta_1){\rm sech}^2\Delta_1,~~\phi=-\frac{c_2}{c_1}-\frac{2c}{c_1}\tanh\Delta_1,\notag
\end{eqnarray}
with $\Delta_1=c(x+\Phi)$. Here $U\equiv U(\xi),V\equiv V(\xi),W\equiv W(\xi),M\equiv M(\xi),F\equiv F(\xi),G\equiv G(\xi),H\equiv H(\xi),$ and $\Phi\equiv \Phi(\xi)$ represent eight different group invariants in (\ref{rmb-22}), while $\xi=-kx+t$  is the similarity variable.

Substituting (\ref{rmb-22}) into the extended system, yields
\begin{eqnarray}\label{rmb-23}
&&U=-\frac{I\mu k\Phi_{\xi\xi}}{k\Phi_{\xi}-1},~~V=-2Ic^2\Phi_{\xi}+2Ikc^2\Phi_{\xi}^2-\frac{Ik^2\Phi_{\xi\xi\xi}}{k\phi_{\xi}-1}+\frac{Ik^3\Phi_{\xi\xi}^2}{(k\phi_{\xi}-1)^2},\notag\\
&&W=-2kc^2\Phi_{\xi}^2+\frac{k^2\Phi_{\xi\xi\xi}+(\mu^2-4c^2)\phi_{\xi}}{k\phi_{\xi}-1}-\frac{k^3\Phi_{\xi\xi}^2+2c^2\phi_{\xi}}{(k\phi_{\xi}-1)^2},\\
&&M=\frac{Ik^2\Phi_{\xi\xi}}{k\phi_{\xi}-1},~~F=\frac{2c^2}{c_1}\Phi_{\xi},~~G=\frac{2kc^2}{c_1}\Phi_{\xi\xi},~~H=\frac{2c^2}{c_1}-\frac{2kc^2}{c_1}\Phi_{\xi},\notag
\end{eqnarray}
where $\Phi$ satisfies the following four-order ordinary differential equation
\begin{equation}\label{rmb-24}
k^2(k\Phi_{\xi}-1)^2\Phi_{\xi\xi\xi\xi}-4k^3(k\Phi_{\xi}-1)\Phi_{\xi\xi\xi}+3k^4\Phi_{\xi\xi}^3-(k\Phi_{\xi}-1)(\mu^2-4c^2+4c^2k^3\Phi_{\xi}^3-12c^2k^2\Phi_{\xi}^2+12c^2k\Phi_{\xi})\Phi_{\xi\xi}=0.
\end{equation}

It is worthy to mention that Eq. (\ref{rmb-24}) is the most general one having the Painlev\'{e} property, that is, having no solutions with movable singularities except poles. According the Ablowitz-Ramani-Segur (ARS) conjecture \cite{j-ablowitz-lnc-1978,j-ablowitz-jmp-1980-1,j-ablowitz-jmp-1980-2}: every nonlinear ordinary differential equation obtained by an exact reduction of a nonlinear partial differential equation of IST class is of Painlev\'{e}-type. By using the ARS algorithm, there is one possible branch: $\Phi(\xi)=A/(\xi-\xi_0)$ with $A=\frac{k}{c^2}$, and the resonant points occur at $\{-1,1,4,6\}$. Then the detailed calculation shows Eq. (\ref{rmb-24}) passes the Painlev\'{e} test and has the Painlev\'{e} property.

It appears that once the solutions $\Phi$ are solved out in (\ref{rmb-24}), the fields for $U,V,W,M,F,G$, and $H$ can be solved out directly from (\ref{rmb-23}). By substituting $\Phi,U,V,W,M,F,G$, and $H$ into (\ref{rmb-22}), the exact solutions of RMB equations (\ref{rmb-03}) are immediately obtained. From the results in reduction 1, one can observe that the final exact solutions include hyperbolic function, Painlev\'{e} wave, and rational function. These solutions represent the wave interactions among solitary wave, Painlev\'{e} wave, and rational wave for the RMB equations. To show the solutions more intuitively, we just take a simple solution of Eq. (\ref{rmb-24}) as $\Phi=m\xi$.

\begin{figure*}[!htbp]
\centering
\subfigure[]{\includegraphics[height=1.4in,width=1.9in]{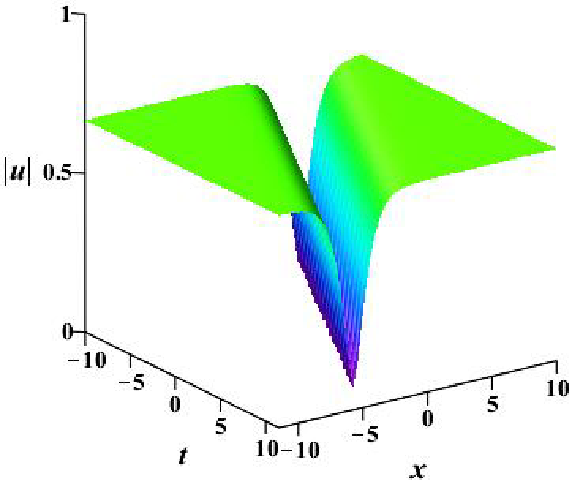}}\hspace{1cm}
\subfigure[]{\includegraphics[height=1.4in,width=1.9in]{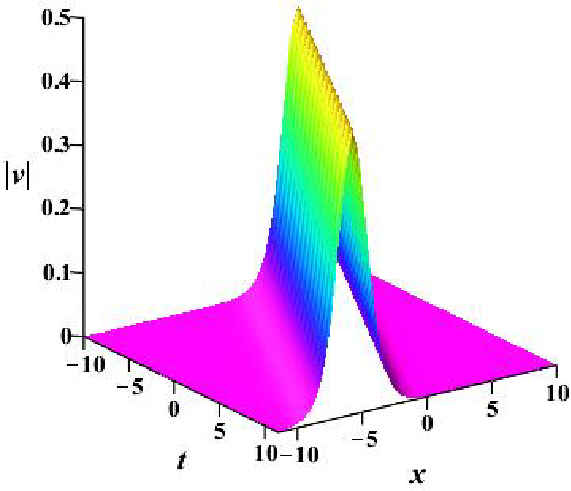}}\hspace{1cm}
\subfigure[]{\includegraphics[height=1.4in,width=1.9in]{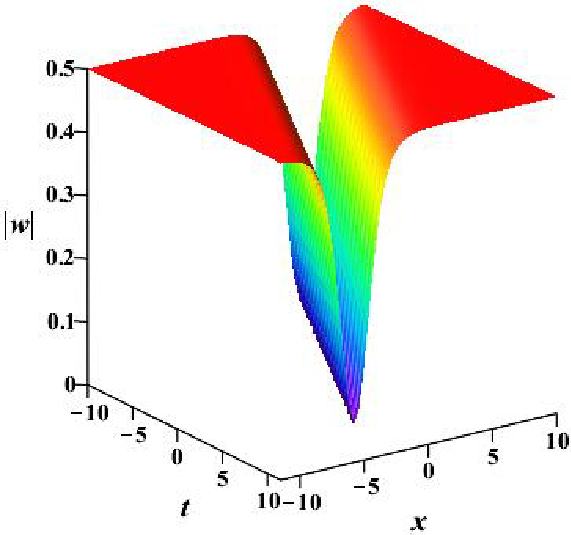}}\hspace{1cm}
\subfigure[]{\includegraphics[height=1.4in,width=1.9in]{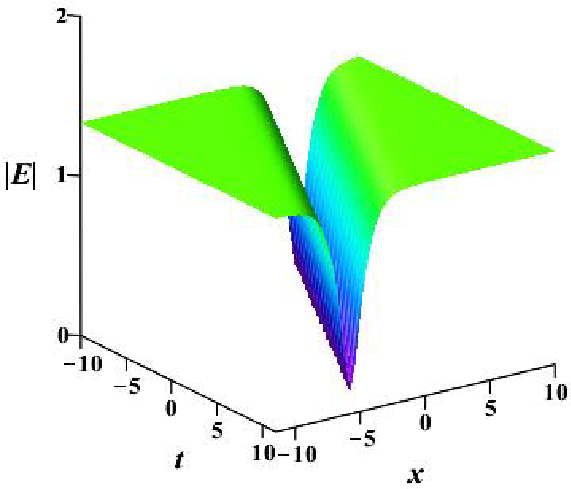}}
\caption{ The wave propagation plots of the RMB equations for the components $u,v,\omega,$ and $E$ expressed by (\ref{rmb-22}). The parameters are $\mu=1,m=\frac{1}{3},k=1,c=1$.}\label{Fig-02}
\end{figure*}

\begin{figure*}[!htbp]
\centering
\subfigure[]{\includegraphics[height=1.4in,width=1.9in]{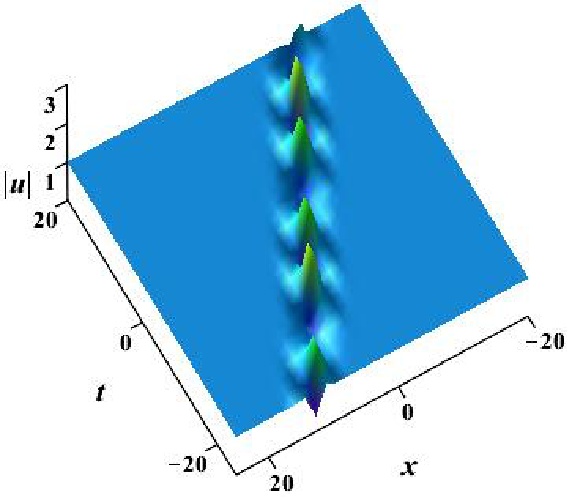}}\hspace{1cm}
\subfigure[]{\includegraphics[height=1.4in,width=1.9in]{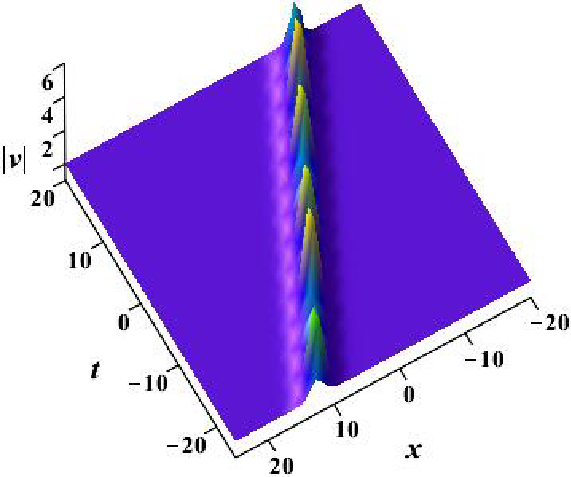}}\hspace{1cm}
\subfigure[]{\includegraphics[height=1.4in,width=1.9in]{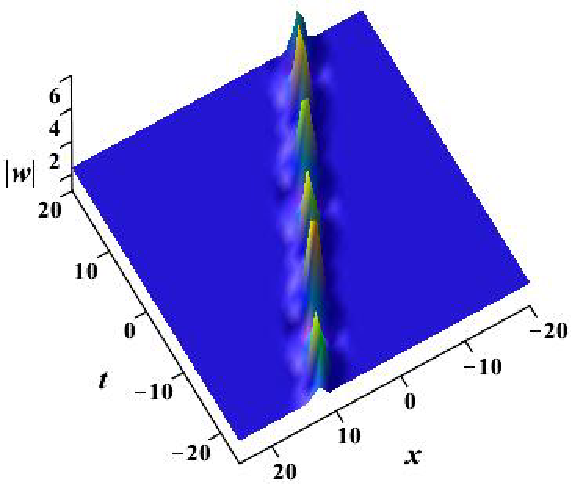}}\hspace{1cm}
\subfigure[]{\includegraphics[height=1.4in,width=1.9in]{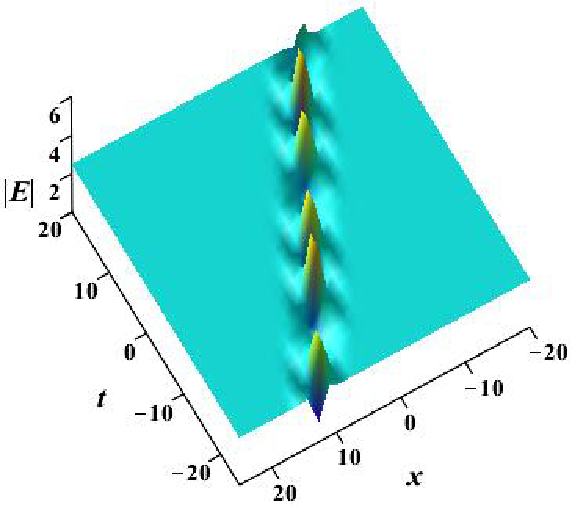}}
\caption{ Breathers of the RMB equations for the components $u,v,\omega,$ and $E$ expressed by (\ref{rmb-22}). The parameters are $\mu=1,m=\frac{1}{2},k=\frac{3}{2}I,c=1$. (a),(b),(c) and (d) The three-dimensional plots for the corresponding solutions.}\label{Fig-03}
\end{figure*}

By choosing the appropriate parameters, different behaviours of $u,v,\omega,$ and $E$ are shown in Fig. (\ref{Fig-02}). It can be seen that $u,\omega,E$ are dark solitary waves, while $v$ is a bright solitary wave. $v$ and $\omega$ have the same amplitudes. If choosing the parameters are $\mu=1,m=\frac{1}{2},k=\frac{3}{2}I,c=1$, we can get breather solutions of the RMB equations. The waves propagation for the components $u,v,\omega,$ and $E$ are shown in Fig. (\ref{Fig-03}), which are different from the Ma breathers in Fig. (\ref{Fig-01-b}).


\textbf{Reduction 2.} Without loss of generality, we hypothesize $f_1=\frac{1}{q_t}$ with $q\equiv q(t)$, and define the parameter $c$ by $c^2=\frac{c_2^2-2c_1c_3}{4}$ $(c\neq0)$. By solving (\ref{rmb-21}), we can derive the similarity solutions
\begin{eqnarray}\label{rmb-25}
&&u=q_t(U+\frac{c_1}{c}I\mu F\tanh\Delta_2),~~v=q_t(V+\frac{c_1}{c}IG\tanh\Delta_2+\frac{c_1^2}{2c^2}IFH\tanh^2\Delta_2),\notag\\
&&\omega=q_t(W-\frac{c_1}{c}G\tanh\Delta_2-\frac{c_1^2}{2c^2}FH\tanh^2\Delta_2),~~E=M+\frac{Ic_1}{c}H\tanh\Delta_2,~~h=-H{\rm sech}^2\Delta_2,\\
&&f=-q_tF{\rm sech}^2\Delta_2,~~g=(G+\frac{c_1}{c}FH\tanh\Delta_2)q_t{\rm sech}^2\Delta_2,~~\phi=-\frac{c_2}{c_1}-\frac{2c}{c_1}\tanh\Delta_2,\notag
\end{eqnarray}
with $\Delta_2=c(q+\Phi)$. Here $U\equiv U(\xi),V\equiv V(\xi),W\equiv W(\xi),M\equiv M(\xi),F\equiv F(\xi),G\equiv G(\xi),H\equiv H(\xi),$ and $\Phi\equiv \Phi(\xi)$ represent eight different group invariants in (\ref{rmb-25}), while $\xi=-c_4q+x$  is the similarity variable.

Substituting (\ref{rmb-25}) into the prolonged system yields
\begin{eqnarray}\label{rmb-26}
&&U=\frac{Ic_4\mu\Phi_{\xi\xi}}{\Phi_{\xi}},~~V=-2Ic^2\Phi_{\xi}+2Ic^2c_4\Phi_{\xi}^2-\frac{Ic_4\Phi_{\xi\xi\xi}}{\phi_{\xi}}+\frac{Ic_4\Phi_{\xi\xi}^2}{\Phi_{\xi}^2},\notag\\
&&W=c_4\mu^2+2c^2\Phi_{\xi}-2c^2c_4\Phi_{\xi}^2+\frac{c_4\Phi_{\xi\xi\xi}-\mu^2}{\Phi_{\xi}}-\frac{c_4\Phi_{\xi\xi}^2}{\Phi_{\xi}^2},\\
&&M=-\frac{I\Phi_{\xi\xi}}{\Phi_{\xi}},~~F=\frac{2c^2}{c_1}-\frac{2c^2c_4}{c_1}\Phi_{\xi},~~G=\frac{2c^2c_4}{c_1}\Phi_{\xi\xi},~~H=\frac{2c^2}{c_1}\Phi_{\xi},\notag
\end{eqnarray}
where $\Phi$ satisfies the following four-order ordinary differential equation
\begin{equation}\label{rmb-27}
3c_4\Phi_{\xi\xi}^3+(\mu^2-4c^2c_4\Phi_{\xi}^3)\Phi_{\xi\xi}-4c_4\Phi_{\xi}\Phi_{\xi\xi}\Phi_{\xi\xi\xi}+c_4\Phi_{\xi}^2\Phi_{\xi\xi\xi\xi}=0.
\end{equation}

By using the ARS algorithm, there is one possible branch: $\Phi(\xi)=A/(\xi-\xi_0)$ with $A=\frac{3}{c^2}$, and the resonant points appear at $\{-1,1,4,6\}$. Then the detailed calculation shows Eq. (\ref{rmb-27}) also passes the Painlev\'{e} test and has the Painlev\'{e} property.

It is also obvious that once the solutions $\Phi$ are solved out in (\ref{rmb-24}), the fields for $U,V,W,M,F,G$, and $H$ can be solved out directly from (\ref{rmb-26}). By substituting $\Phi,U,V,W,M,F,G$, and $H$ into (\ref{rmb-25}), the exact solutions of RMB equations (\ref{rmb-03}) are immediately obtained. From the results in reduction 2, one can observe that the final exact solutions include hyperbolic function, Painlev\'{e} wave, and rational function. These solutions represent the waves interaction among solitary waves, Painlev\'{e} waves, and rational waves for the RMB equations.

Because of the existence of the arbitrary function $q$ in (\ref{rmb-25}), many new types of interaction solutions can be obtained. Since $q$ can be expressed by different types of functions, the solutions exhibit the interactions between solitary waves and abundant cnoidal waves, rational waves, trigonometric functions. In the process of interaction, many localized excitations including rogue waves and breathers are appeared. To show these localized excitation states more intuitively, we just take a simple solution of Eq. (\ref{rmb-27}) as $\Phi=m\xi$.


\begin{figure*}[!htbp]
\centering
\subfigure[]{\includegraphics[height=1.1in,width=1.4in]{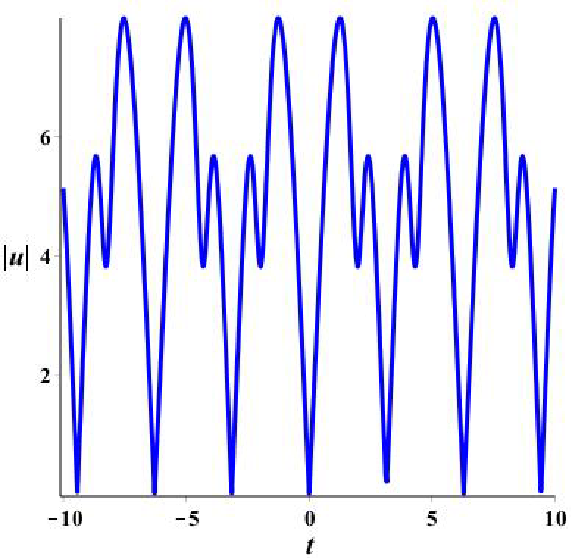}}\hspace{0.1cm}
\subfigure[]{\includegraphics[height=1.1in,width=1.4in]{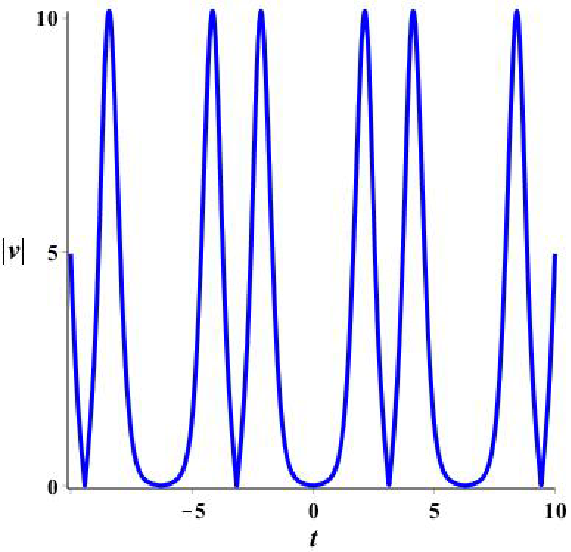}}\hspace{0.1cm}
\subfigure[]{\includegraphics[height=1.1in,width=1.4in]{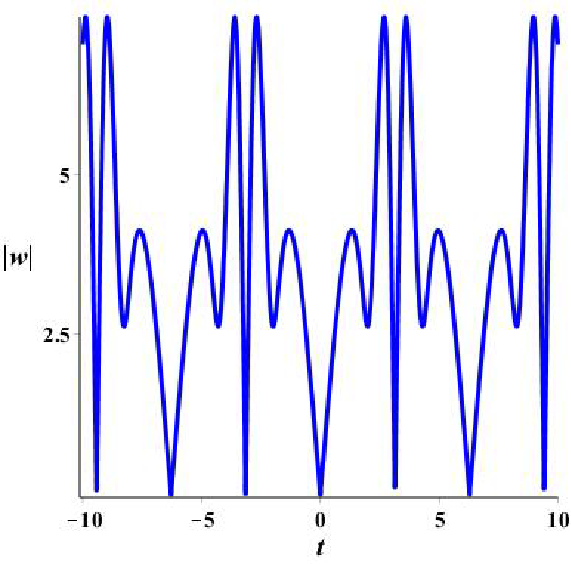}}\hspace{0.1cm}
\subfigure[]{\includegraphics[height=1.1in,width=1.4in]{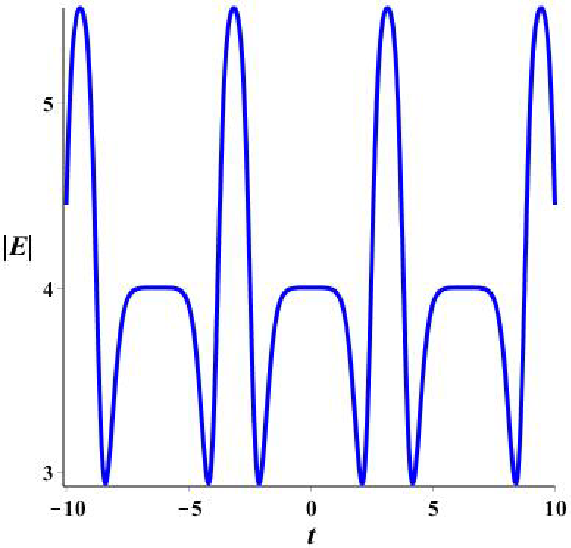}}\hspace{0.1cm}
\subfigure[]{\includegraphics[height=1.1in,width=1.4in]{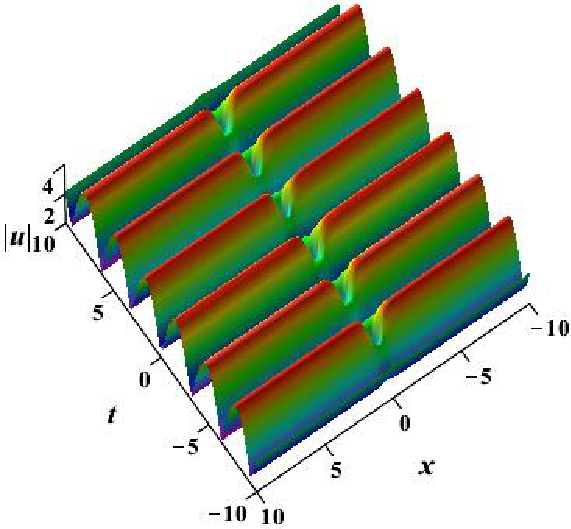}}\hspace{0.1cm}
\subfigure[]{\includegraphics[height=1.1in,width=1.4in]{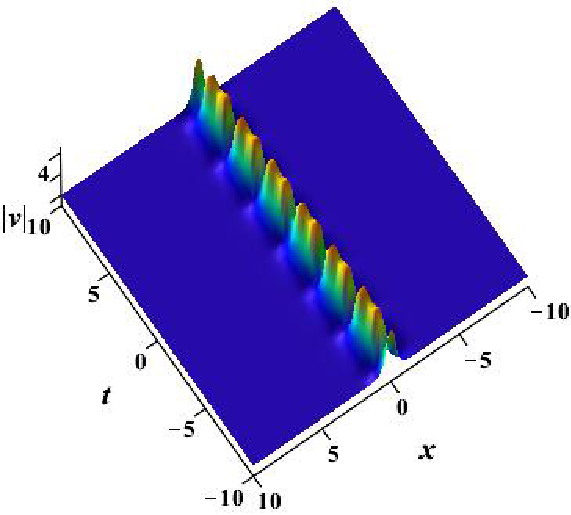}}\hspace{0.1cm}
\subfigure[]{\includegraphics[height=1.1in,width=1.4in]{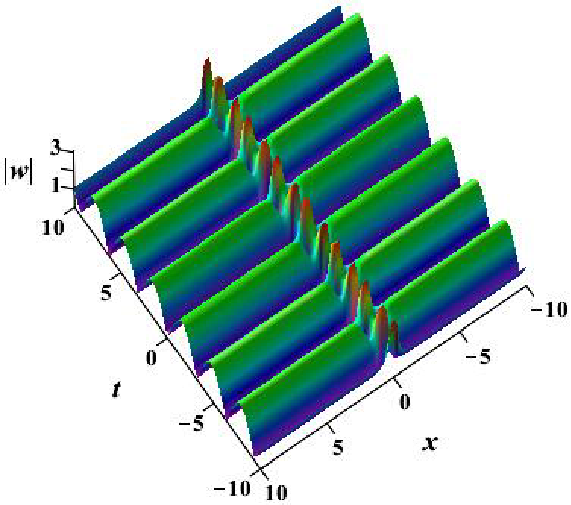}}\hspace{0.1cm}
\subfigure[]{\includegraphics[height=1.1in,width=1.4in]{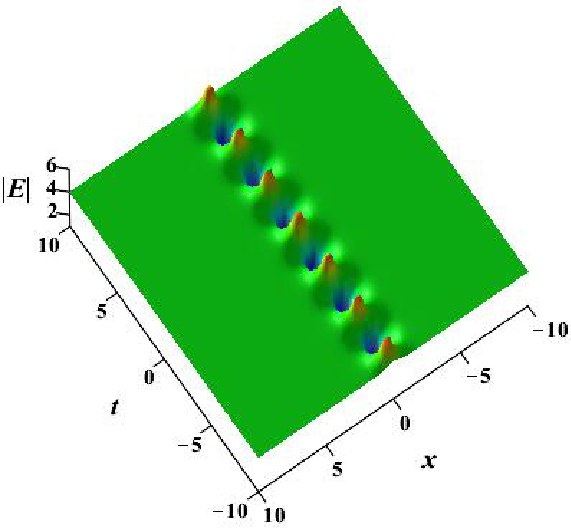}}
\caption{ The periodic wave solutions of the RMB equations for the components $u,v,\omega,$ and $E$ expressed by (\ref{rmb-25}). The parameters are $\mu=2,m=2,c_4=I,c=1,q=I\cos(t)$. (a),(b),(c) and (d) The waves propagation along $t$ axis at $x=0.5$; (e),(f),(g) and (h) The three-dimensional plots for the corresponding solutions.}\label{Fig-04}
\end{figure*}

\begin{figure*}[!htbp]
\centering
\subfigure[]{\includegraphics[height=1.4in,width=1.9in]{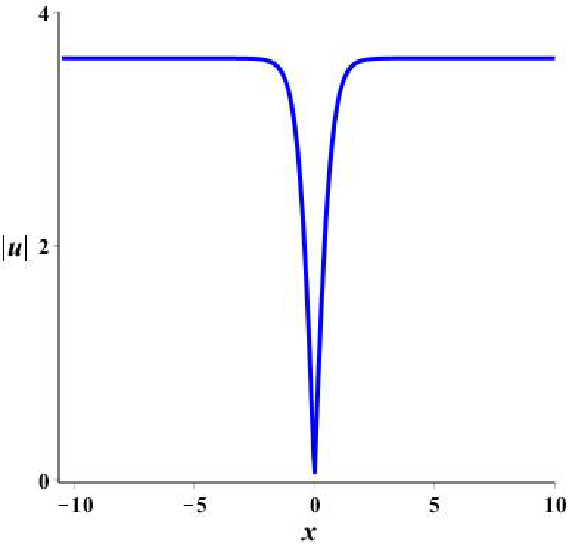}}\hspace{0.5cm}
\subfigure[]{\includegraphics[height=1.4in,width=1.9in]{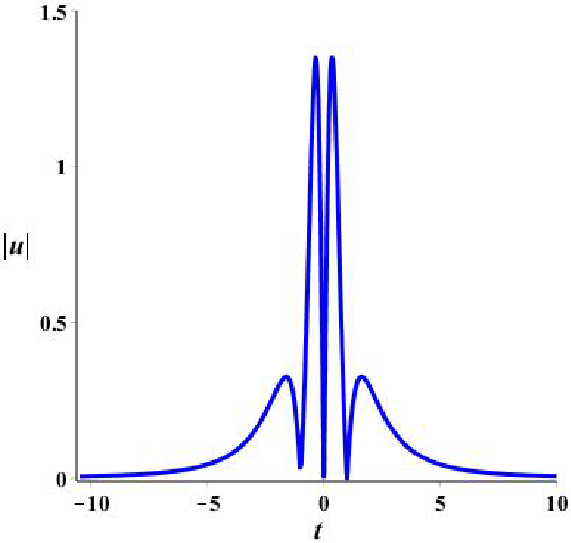}}\hspace{0.5cm}
\subfigure[]{\includegraphics[height=1.4in,width=1.9in]{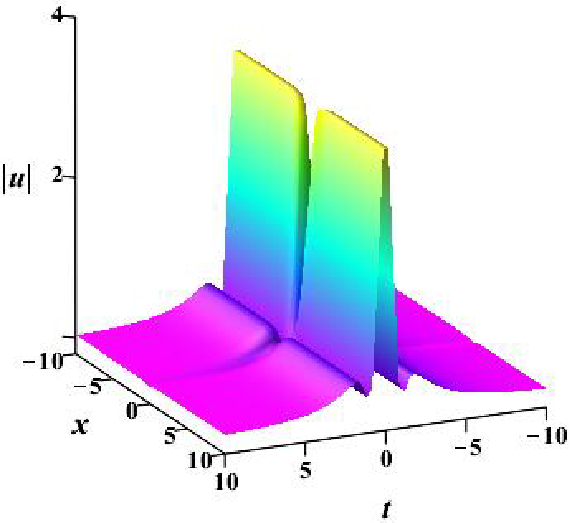}}\hspace{0.5cm}
\subfigure[]{\includegraphics[height=1.4in,width=1.9in]{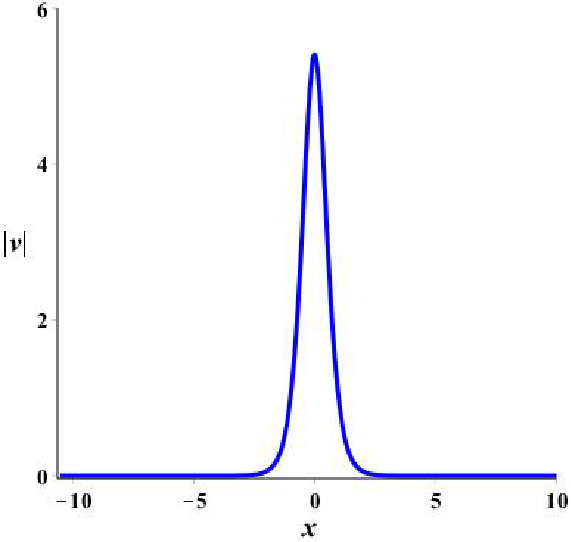}}\hspace{0.5cm}
\subfigure[]{\includegraphics[height=1.4in,width=1.9in]{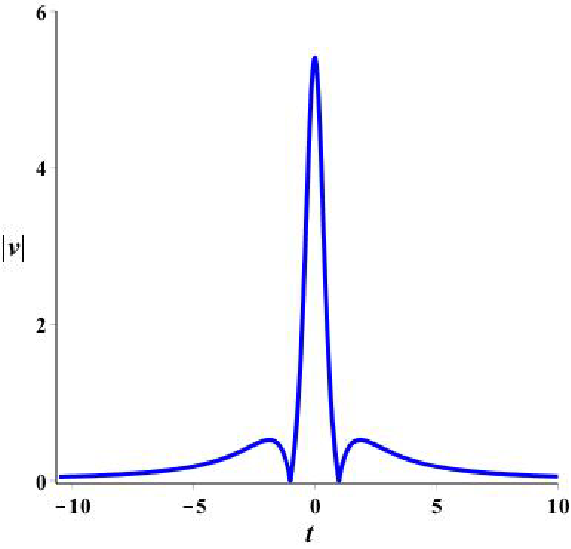}}\hspace{0.5cm}
\subfigure[]{\includegraphics[height=1.4in,width=1.9in]{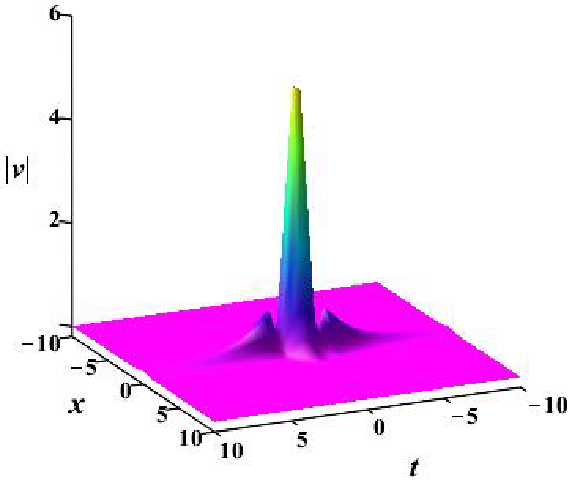}}\hspace{0.5cm}
\subfigure[]{\includegraphics[height=1.4in,width=1.9in]{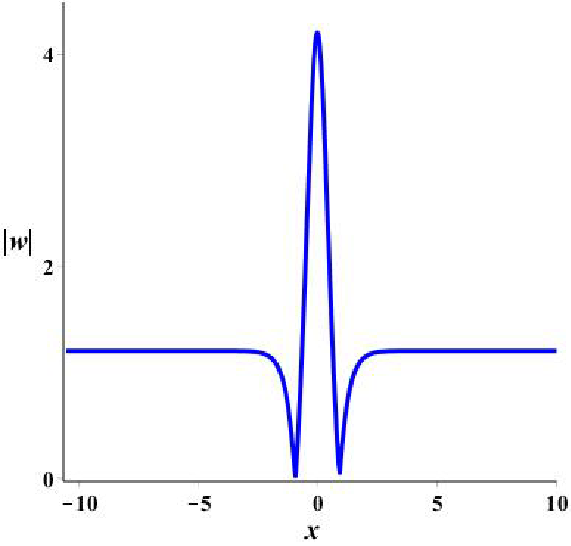}}\hspace{0.5cm}
\subfigure[]{\includegraphics[height=1.4in,width=1.9in]{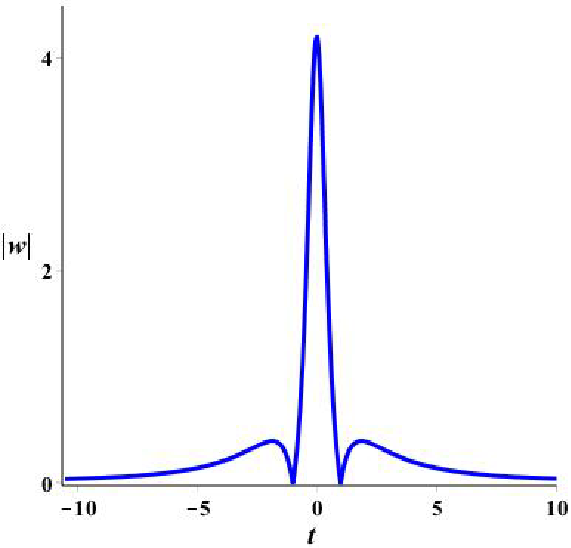}}\hspace{0.5cm}
\subfigure[]{\includegraphics[height=1.4in,width=1.9in]{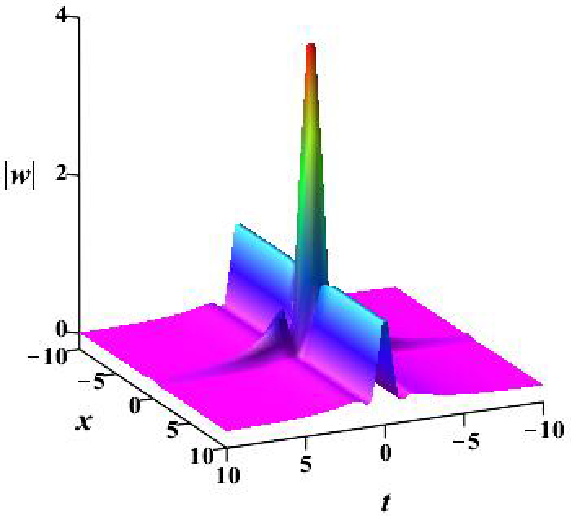}}\hspace{0.5cm}
\subfigure[]{\includegraphics[height=1.4in,width=1.9in]{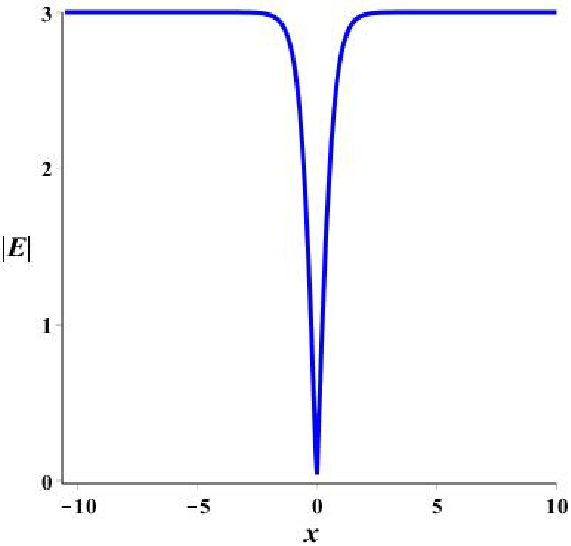}}\hspace{0.5cm}
\subfigure[]{\includegraphics[height=1.4in,width=1.9in]{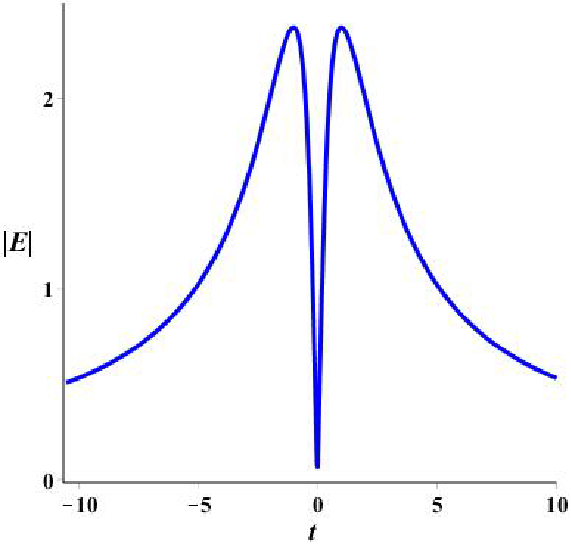}}\hspace{0.5cm}
\subfigure[]{\includegraphics[height=1.4in,width=1.9in]{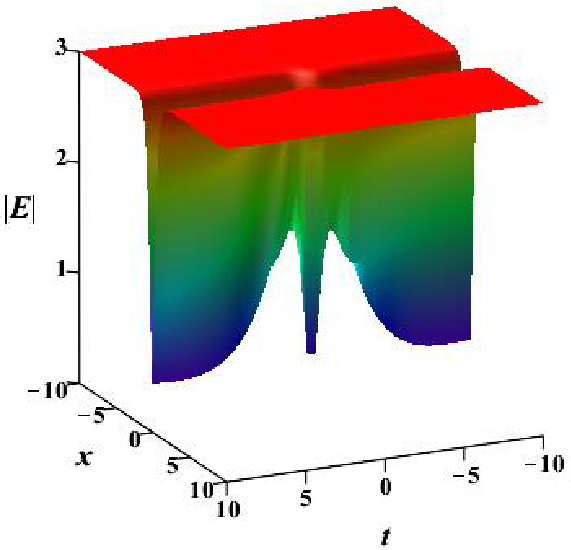}}
\caption{ The localized excitations of the RMB equations for the components $u,v,\omega,$ and $E$ expressed by (\ref{rmb-25}). The parameters are $\mu=1,m=\frac{3}{2},c_4=I,c=1,q=\frac{It}{1+t^2}$. (a),(d),(g) and (j) The waves propagate along $x$ axis at $x=0$; (b),(e), and (h) The waves propagate along $t$ axis at $t=0$; (k) The wave propagates along $t$ axis at $t=0.1$; (c),(f),(i) and (l) The three-dimensional plots for the corresponding solutions.}\label{Fig-05}
\end{figure*}

\begin{figure*}[!htbp]
\centering
\subfigure[]{\includegraphics[height=1.4in,width=1.9in]{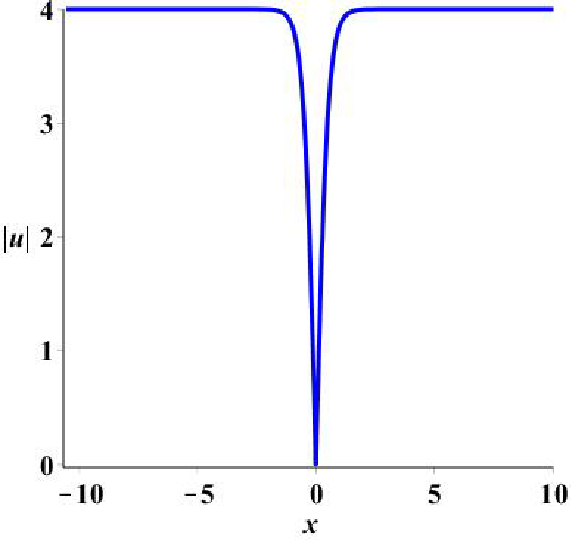}}\hspace{0.5cm}
\subfigure[]{\includegraphics[height=1.4in,width=1.9in]{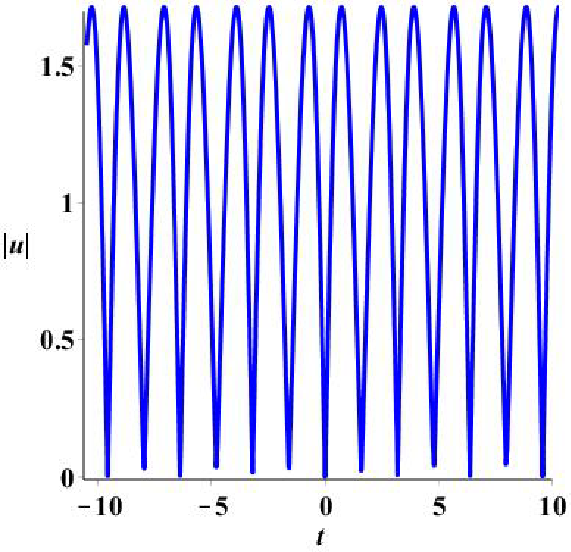}}\hspace{0.5cm}
\subfigure[]{\includegraphics[height=1.4in,width=1.9in]{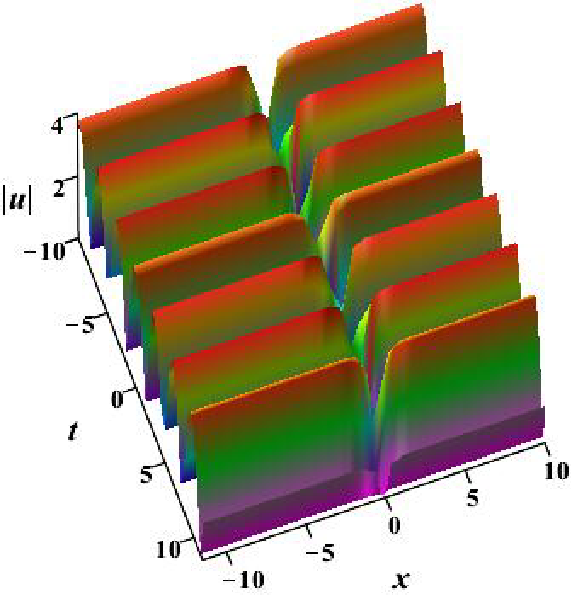}}\hspace{0.5cm}
\subfigure[]{\includegraphics[height=1.4in,width=1.9in]{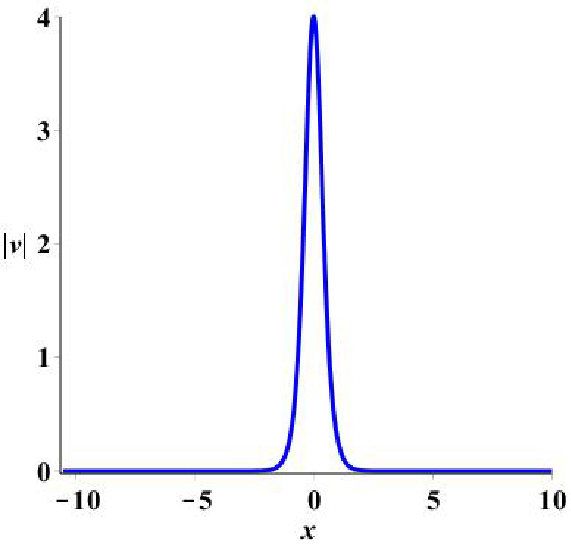}}\hspace{0.5cm}
\subfigure[]{\includegraphics[height=1.4in,width=1.9in]{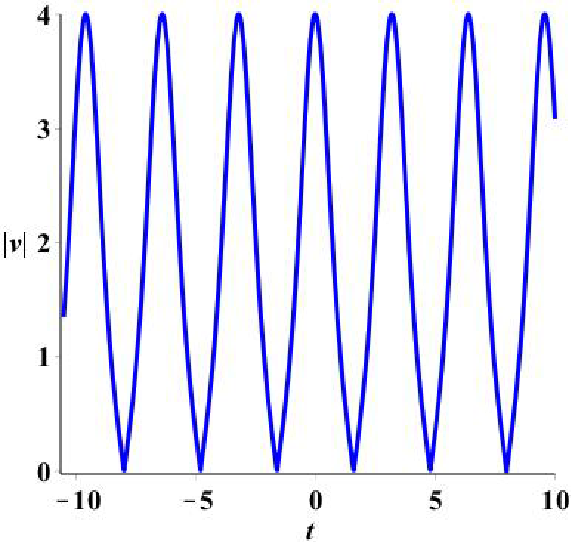}}\hspace{0.5cm}
\subfigure[]{\includegraphics[height=1.4in,width=1.9in]{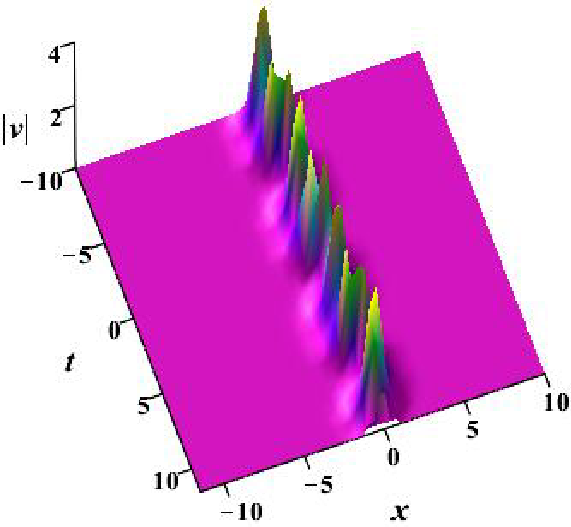}}\hspace{0.5cm}
\subfigure[]{\includegraphics[height=1.4in,width=1.9in]{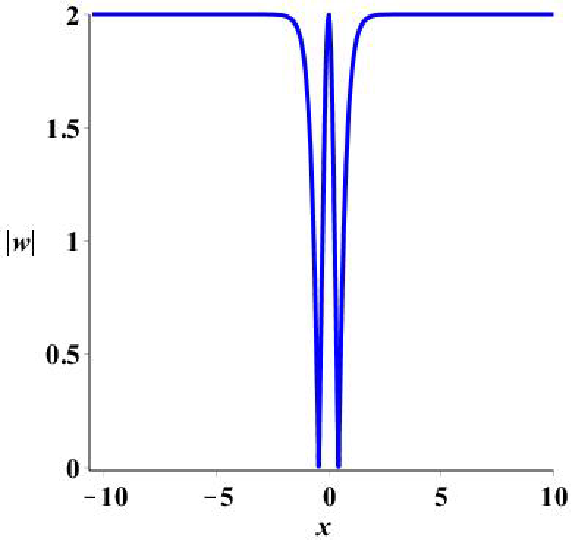}}\hspace{0.5cm}
\subfigure[]{\includegraphics[height=1.4in,width=1.9in]{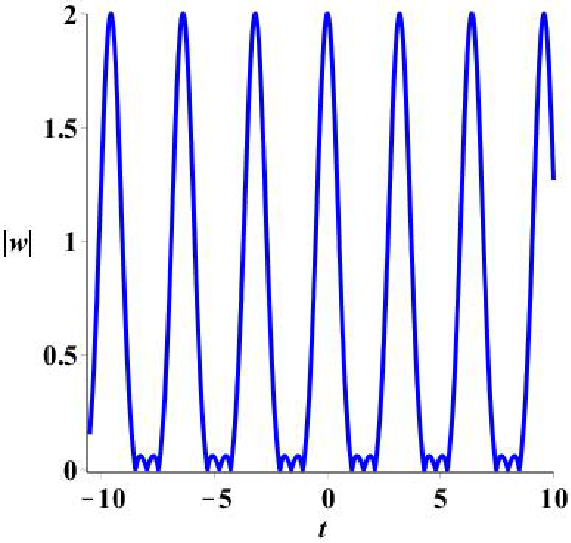}}\hspace{0.5cm}
\subfigure[]{\includegraphics[height=1.4in,width=1.9in]{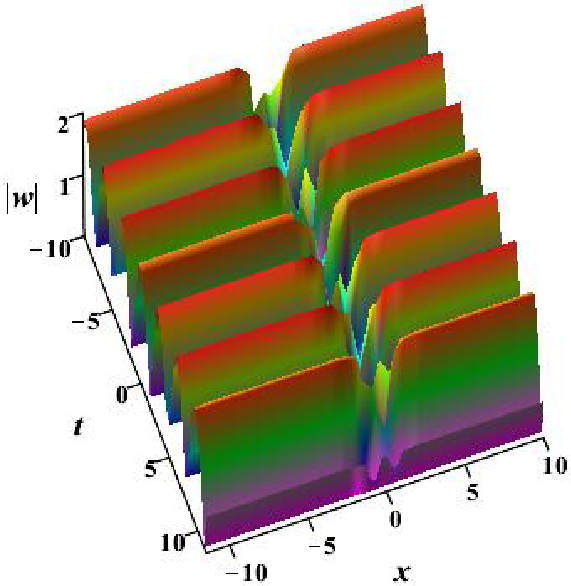}}\hspace{0.5cm}
\subfigure[]{\includegraphics[height=1.4in,width=1.9in]{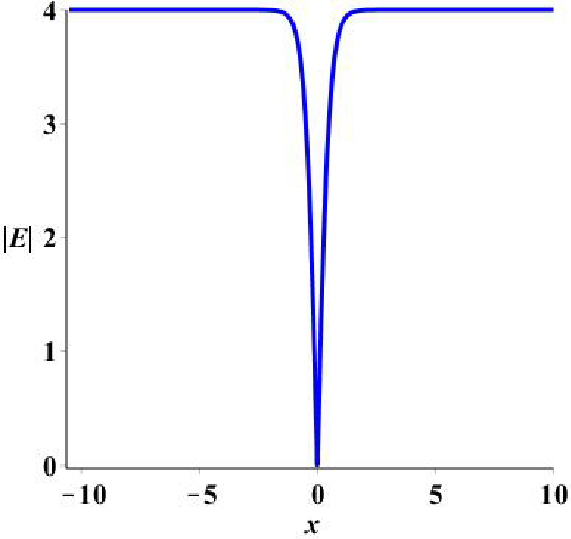}}\hspace{0.5cm}
\subfigure[]{\includegraphics[height=1.4in,width=1.9in]{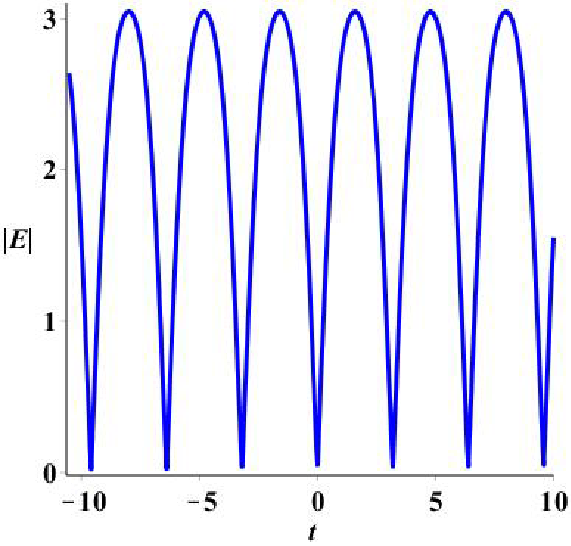}}\hspace{0.5cm}
\subfigure[]{\includegraphics[height=1.4in,width=1.9in]{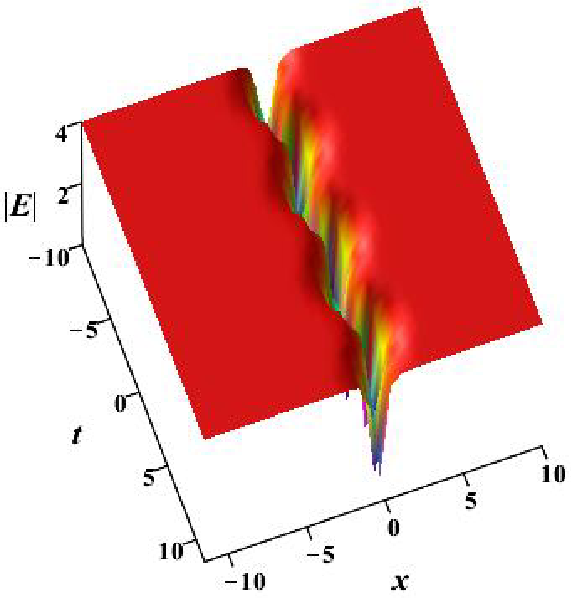}}
\caption{ The wave propagation plots of the RMB equations for solutions $u,v,\omega,$ and $E$ expressed by (\ref{rmb-25}). The parameters are $\mu=2,m=2,n=\frac{1}{4},c_4=1,c=1,q=sn(t,n)$. (a),(d),(g) and (j) The wave propagation pattern of the wave along $x$ axis; (b),(e),(h) and (k) The wave propagation pattern of the wave along $t$ axis; (c),(f),(i) and (l) The three-dimensional plots for the corresponding solutions. }\label{Fig-06}
\end{figure*}

\begin{figure*}[!htbp]
\centering
\subfigure[]{\includegraphics[height=1.4in,width=1.9in]{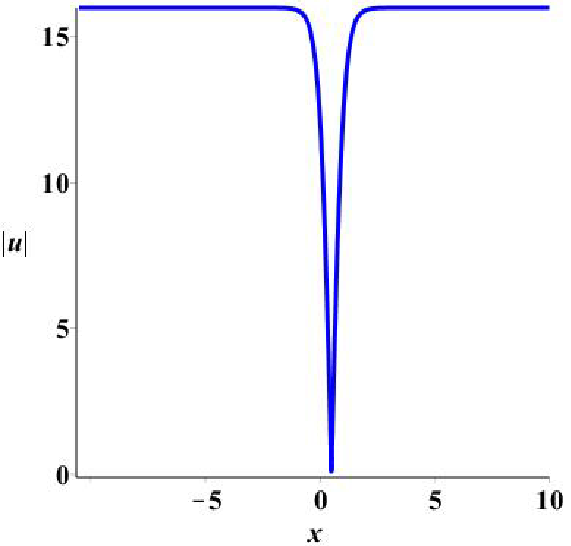}}\hspace{0.5cm}
\subfigure[]{\includegraphics[height=1.4in,width=1.9in]{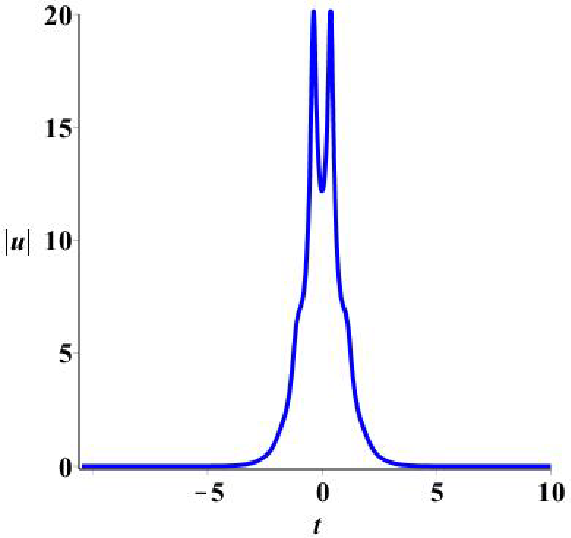}}\hspace{0.5cm}
\subfigure[]{\includegraphics[height=1.4in,width=1.9in]{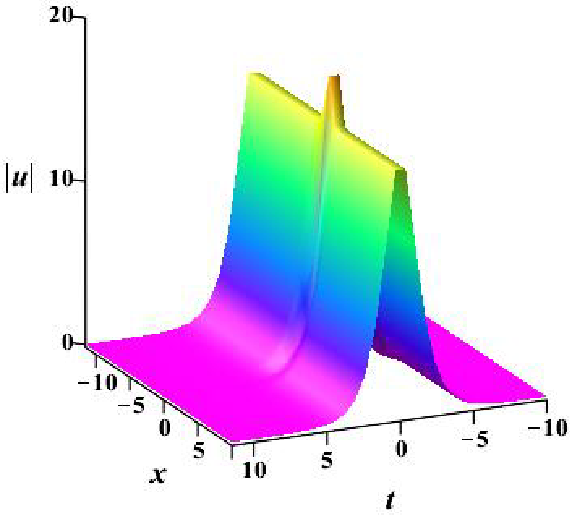}}\hspace{0.5cm}
\subfigure[]{\includegraphics[height=1.4in,width=1.9in]{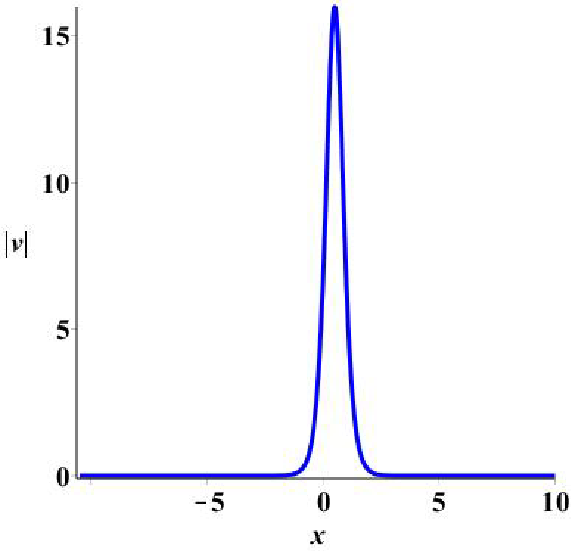}}\hspace{0.5cm}
\subfigure[]{\includegraphics[height=1.4in,width=1.9in]{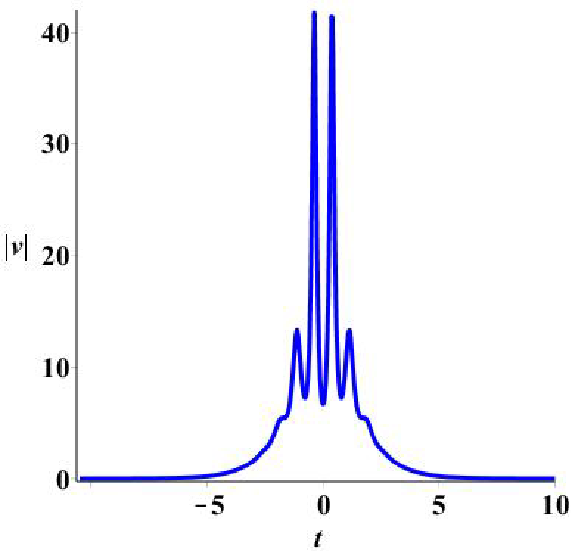}}\hspace{0.5cm}
\subfigure[]{\includegraphics[height=1.4in,width=1.9in]{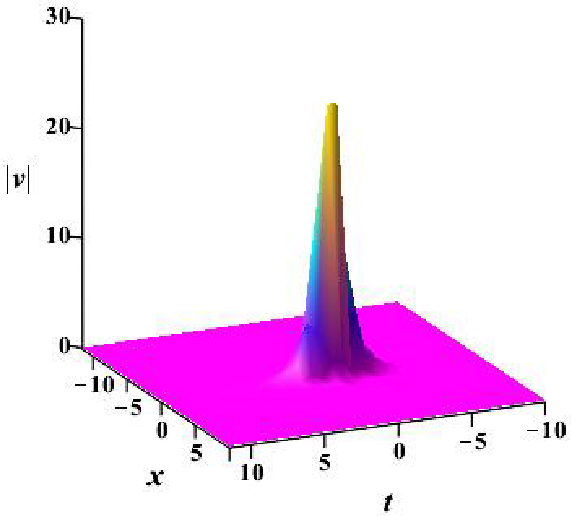}}\hspace{0.5cm}
\subfigure[]{\includegraphics[height=1.4in,width=1.9in]{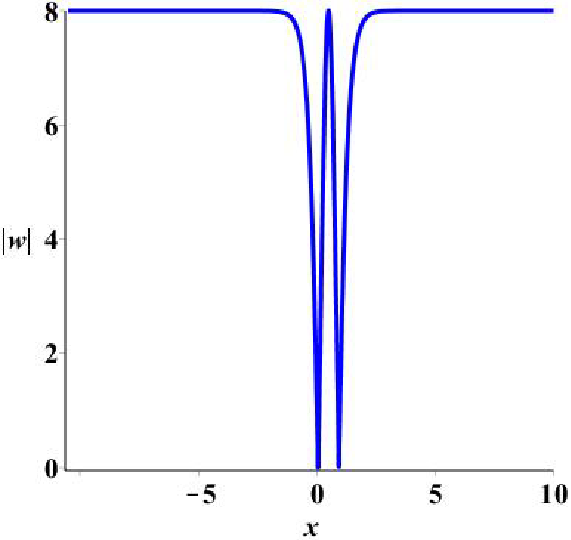}}\hspace{0.5cm}
\subfigure[]{\includegraphics[height=1.4in,width=1.9in]{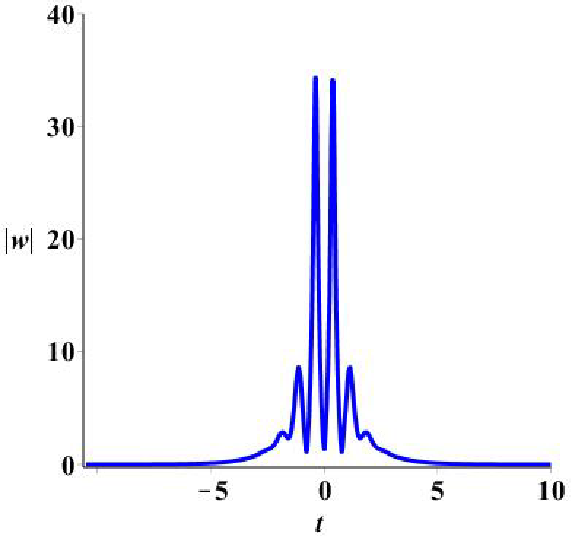}}\hspace{0.5cm}
\subfigure[]{\includegraphics[height=1.4in,width=1.9in]{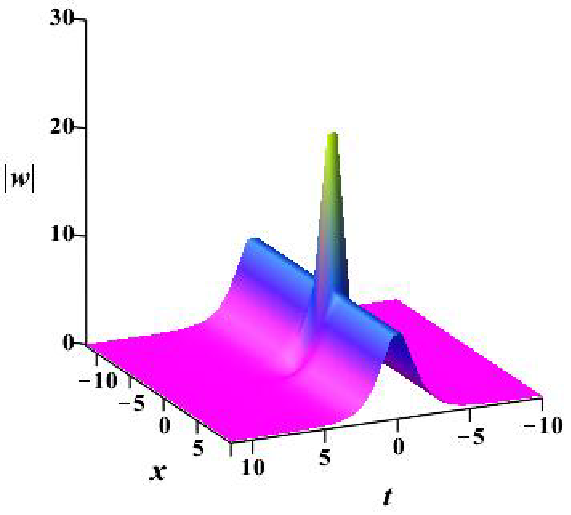}}\hspace{0.5cm}
\subfigure[]{\includegraphics[height=1.4in,width=1.9in]{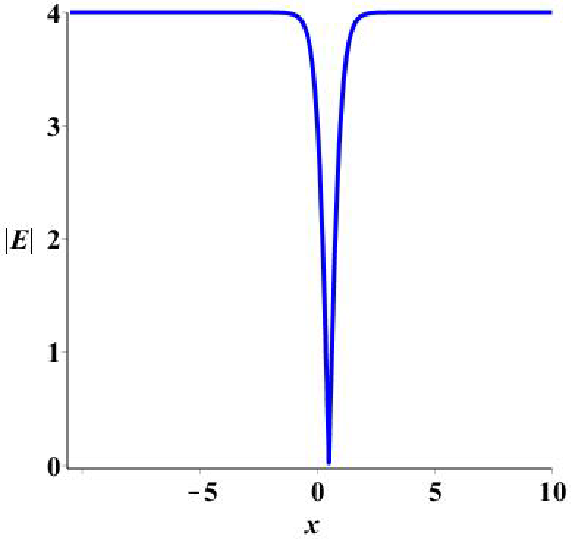}}\hspace{0.5cm}
\subfigure[]{\includegraphics[height=1.4in,width=1.9in]{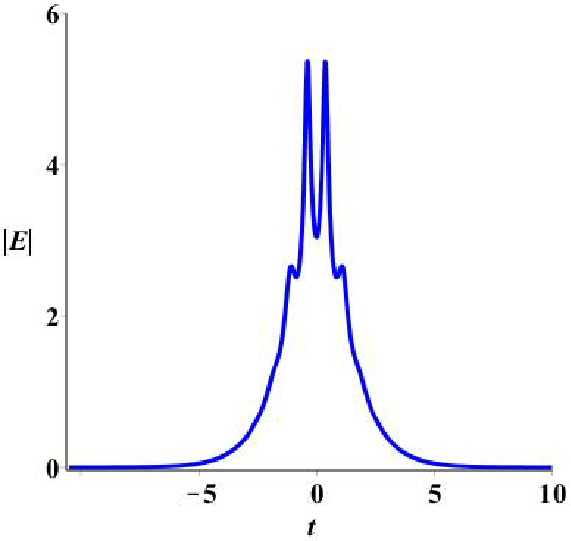}}\hspace{0.5cm}
\subfigure[]{\includegraphics[height=1.4in,width=1.9in]{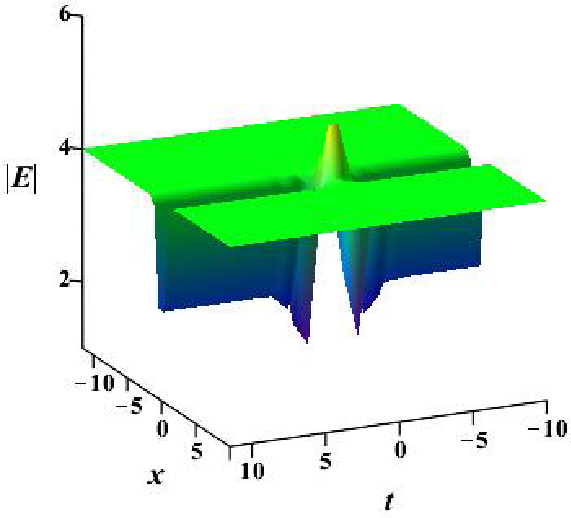}}
\caption{ The localized excitations of the RMB equations for the components $u,v,\omega,$ and $E$ expressed by (\ref{rmb-25}). The parameters are $\mu=2,m=2,c_4=1,c=1,q=\exp^{4It}sech(t)$. (a),(d),(g) and (j) The wave propagation pattern of the wave along $x$ axis; (b),(e),(h) and (k) The wave propagation pattern of the wave along $t$ axis; (c),(f),(i) and (l) The three-dimensional plots for the corresponding solutions. }\label{Fig-08}
\end{figure*}

\begin{figure*}[!htbp]
\centering
\subfigure[]{\includegraphics[height=1.1in,width=1.4in]{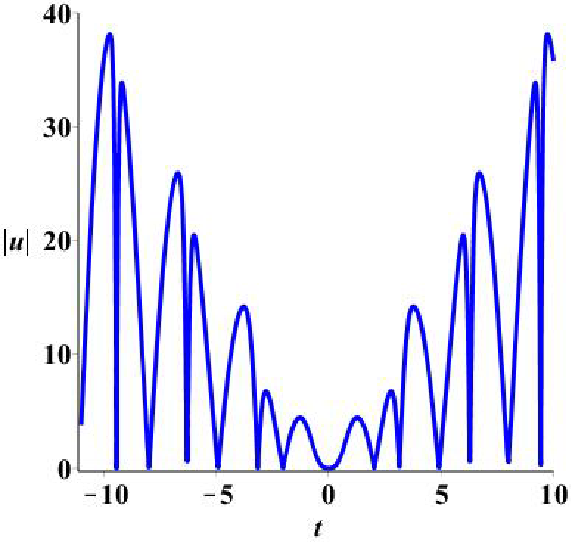}}\hspace{0.1cm}
\subfigure[]{\includegraphics[height=1.1in,width=1.4in]{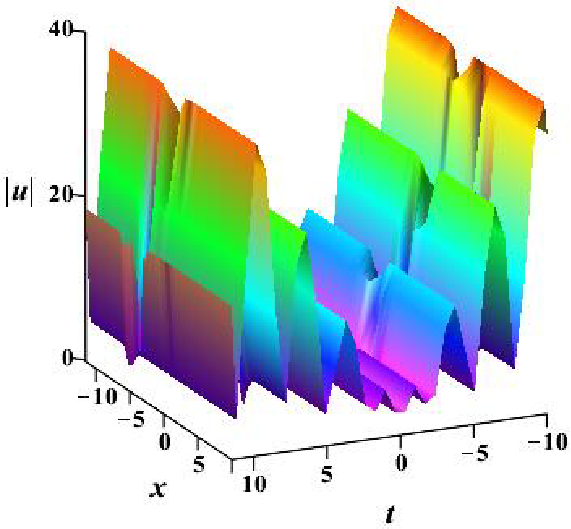}}\hspace{0.1cm}
\subfigure[]{\includegraphics[height=1.1in,width=1.4in]{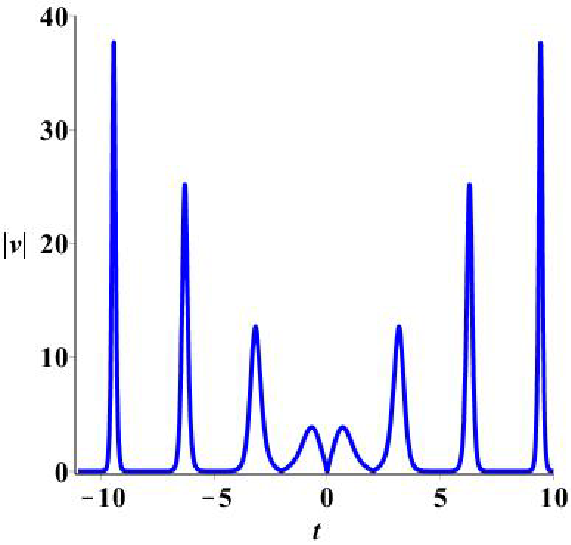}}\hspace{0.1cm}
\subfigure[]{\includegraphics[height=1.1in,width=1.4in]{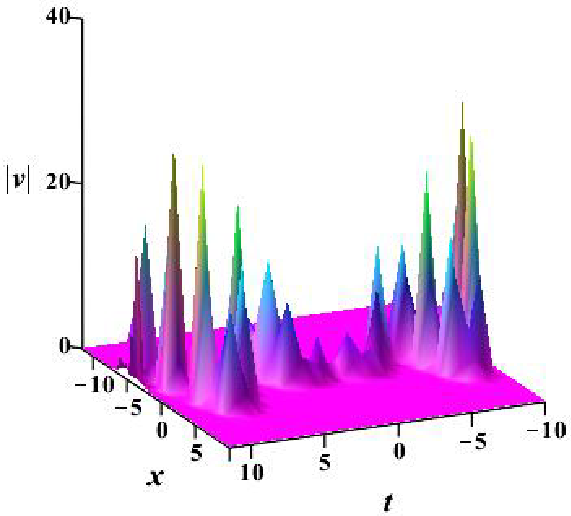}}\hspace{0.1cm}
\subfigure[]{\includegraphics[height=1.1in,width=1.4in]{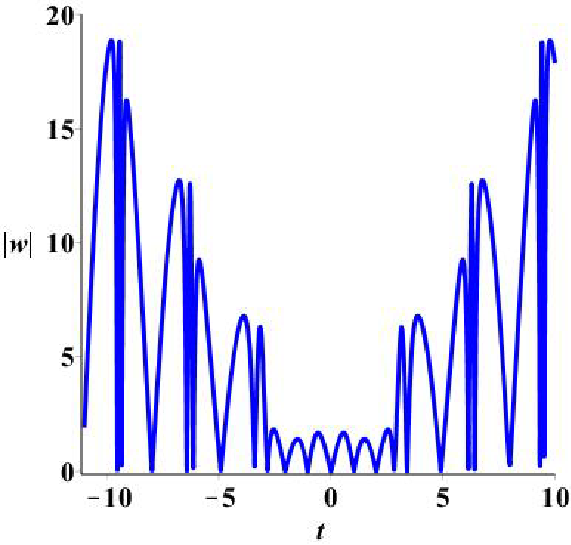}}\hspace{0.1cm}
\subfigure[]{\includegraphics[height=1.1in,width=1.4in]{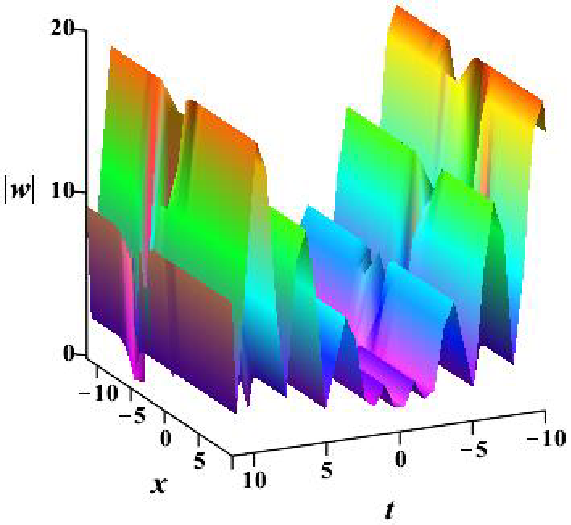}}\hspace{0.1cm}
\subfigure[]{\includegraphics[height=1.1in,width=1.4in]{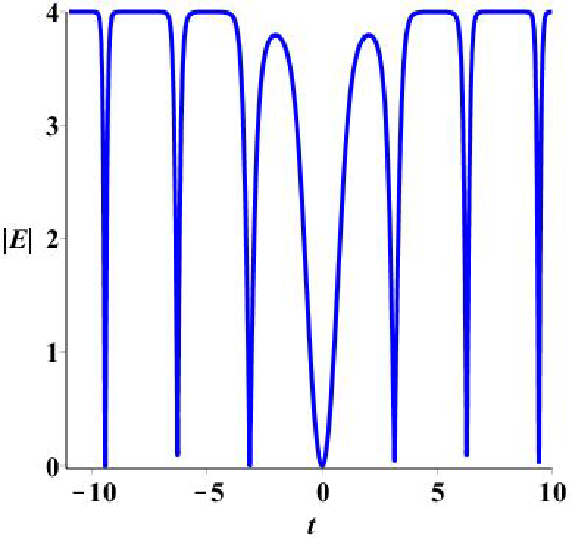}}\hspace{0.1cm}
\subfigure[]{\includegraphics[height=1.1in,width=1.4in]{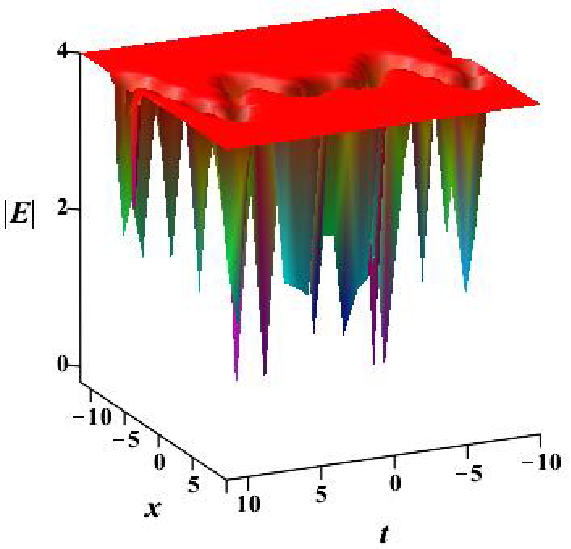}}
\caption{ The wave propagation plots of the RMB equations for the components $u,v,\omega,$ and $E$ expressed by (\ref{rmb-25}). The parameters are $\mu=2,m=2,c_4=1,c=1,q=t\sin(t)$. (a),(c),(e) and (g) The wave propagation pattern of the wave along $t$ axis; (b),(d),(f) and (h) The three-dimensional plots for the corresponding solutions.}\label{Fig-10}
\end{figure*}


$\bullet$~~For Fig. (\ref{Fig-04}), the solutions of the RMB equations for the components $u,v,\omega,$ and $E$ are hybrid solutions with the parameters $\mu=2,m=2,c_4=I,c=1$ and $q=I\cos(t)$. It can be seen that breathers travelling on the background of periodic waves for the components $u$ and $\omega$. Both the breathers and periodic waves are periodic in $t$ directions and localized in $x$ directions. The solutions for the components $v$ and $E$ describe breathers in $t$ directions. These phenomena are similar with Fig. (\ref{Fig-02-b}).

$\bullet$~~For Fig. (\ref{Fig-05}), the localized excitation solutions of the RMB equations for the components $u,v,\omega,$ and $E$ are drawn with the parameters $\mu=1,m=\frac{3}{2},c_4=I,c=1$ and $q=\frac{It}{1+t^2}$. For the component $u$, it changes from a single peak to the double peak at $x=0$, and the amplitude becomes smaller. For the component $v$, a W-type solitary wave appears at $t=0$ and disappears at infinity along $x$ axis. For the component $\omega$, a W-type solitary wave unchanges along $x$ axis, but the amplitude of peak gets smaller at infinity along $t$ axis. For the component $E$, an anti-W-type solitary wave appears at $t=0$ and disappears to the plane wave $E=3$ at infinity along $x$ axis.

$\bullet$~~For Fig. (\ref{Fig-06}), the parameters are $\mu=2,m=2,n=\frac{1}{4},c_4=1,c=1$, and $q=sn(t,n)$. The exact solutions (\ref{rmb-25}) with $\Phi=m\xi$ for the RMB equations denote the interactional phenomena between solitary waves and cnoidal periodic waves. These kinds of solutions can be easily applicable to the analysis of interesting physical phenomena. In fact, there are full of the solitary waves and the cnoidal periodic waves in the real physics world.

$\bullet$~~If the parameters are $\mu=2,m=2,c_4=1,c=1$, and $q=\tanh(t)$. For the component $u$, a dark solitary wave appears at $t=0$ and disappears at infinity along $x$ axis; it changes from a single peak to the double peak at $x=0$, and the amplitude becomes smaller at infinity along $t$ axis. For the component $v$, the wave becomes a bright solitary wave both at $t=0$ and $x=0$, disappears at infinity along $x$ axis and $t$ axis. For the component $\omega$, a W-type solitary wave appears at $t=0$ and disappears at infinity along $x$ axis; the waveform is narrowed at $x=0$ and the amplitude is constant along $t$ axis. For the component $E$, a kink solitary wave appears at $x=0$, the waveform and amplitude unchanges over time, but the wave propagates to the right at $\mid t\mid<41$.

$\bullet$~~For Fig. (\ref{Fig-08}), the parameters are $\mu=2,m=2,c_4=1,c=1$, and $q=\exp(4It)sech(t)$. For the component $u$, a single peak solitary wave appears at $\mid t\mid\leq333$, it becomes a double peak rogue wave at $x=0$, and when $\mid t\mid>333$, the wave disappears along $t$ axis. For the component $v$, a double peak rogue wave appears at $x=0$ and disappears at infinity along $t$ axis. For the component $\omega$, a single peak solitary wave appears at $\mid t\mid\leq333$, it becomes a double peak rogue wave at $x=0$, and when $\mid t\mid>333$, the wave disappears along $t$ axis. For the component $E$, the amplitude unchanges over time, it becomes a double peak rogue wave at $x=0$. but the wave first propagates to the right and then to the left in $\mid t\mid<41$.

$\bullet$~~For Fig. (\ref{Fig-10}), the parameters are $\mu=2,m=2,c_4=1,c=1$, and $q=t\sin(t)$. For the components $u,v,\omega,$ and $E$, the waves generate in pairs, and the solutions are symmetrical at $t=0$.

\textbf{Remark}: Due to $q$ is the function of $t$, the dynamic behaviours of $u,v,\omega,$ and $E$ at $t=0$ are the same and their modules are the same in Figs. (\ref{Fig-06}) and (\ref{Fig-08}).

\section{Summary and discussions}

In summary, nonlocal symmetry, localized excitations and interactional solutions of the RMB equations have been investigated. Based on the truncated Painlev\'{e} expansion approach, nonlocal symmetries of the RMB equations are obtained. And the Schwartzian form of the RMB equations is reduced from the truncated Painlev\'{e} expansion. Meanwhile, the nonlocal symmetries are related to the M\"{o}bious transformation of the Schwartzian form. By introducing three auxiliary variables, the nonlocal symmetries are successfully localized to an extended system, which is closed to a Lie point symmetry and a new type of finite symmetry transformations is derived by solving the initial value problems. Then we obtain the following results, which is difficult to find stemming from nonconstant nonlinear wave such as the cnoidal waves and Painlev\'{e} waves.

$\bullet$ Periodic solutions. Based on the finite symmetry transformations, we obtain three types of solutions including periodic waves (see Fig. \ref{Fig-01}), Ma breathers (see Fig. \ref{Fig-01-b}) and breathers travelling on the background of periodic line waves (see Fig. \ref{Fig-02-b}). The periodic waves are periodic in both $x$ directions and $t$ directions. The breathers are periodic in $t$ directions and localized in $x$ directions.

$\bullet$ Interactional solutions. By symmetry reduction method, rich exact interactional solutions are derived between solitary waves and other waves, such as cnoidal waves, rational solutions, Painlev\'{e} waves, and periodic waves (see Figs. \ref{Fig-04},~\ref{Fig-06}).

$\bullet$ Localized excitations. To our knowledge, the rogue wave solution of the integrable system has not yet been found by the nonlocal symmetry method. Because of arbitrary function generated during the similarity reduction process, several new types of localized excitations including rogue waves, breathers and other nonlinear waves are obtained (see Figs. \ref{Fig-05},~\ref{Fig-08}). Some interesting dynamical behaviour are shown by selecting different functions in graphical way. The rogue wave obtained in this paper may be called instanton or ghost soliton better.

Due to the important physics significance of the RMB equations, the physical properties of those new dynamical behaviour with interactions are needed to do a further investigation. These kinds of solutions can be applied to the analysis of many interesting physical phenomena and may be useful for studying the optical waves and electromagnetic waves. Our research results may play a significant contribution to investigate the dynamical properties of the distinct nonlinear waves, such as rogue waves, breathers, dark solitons and bright solitons, for the nonlinear systems in optics, electromagnetic field, plasma physics and Bose-Einstein condensates. The results may motivate the relevant experimental investigations in the ultra-short optical pulses and other media.

\section*{Acknowledgment}
We would like to express our sincere thanks to other members of our discussion group for their valuable suggestions.
The project is supported by the Global Change Research
Program of China (No. 2015CB953904), the National Natural Science Foundation of China (Nos. 11675054 and 11435005), Outstanding doctoral dissertation cultivation plan of action (No. YB2016039), and Shanghai Collaborative Innovation Center of
Trustworthy Software for Internet of Things (No. ZF1213).

\end{document}